\documentclass[pre,amsmath,twocolumn,showpacs]{revtex4}
\usepackage{bm}
\usepackage{amssymb}
\usepackage{graphicx}
\usepackage{subfigure}

\def\r{{\mbox{\boldmath{$\rho$}}}}

\begin{document}

\title{Jamming transition of kinetically-constrained models in rectangular systems}

\author{Eial Teomy}
\email{eialteom@post.tau.ac.il}
\author{Yair Shokef}
\email{shokef@tau.ac.il}

\affiliation{School of Mechanical Engineering, Tel Aviv University, Tel Aviv 69978, Israel}

\begin{abstract}
We theoretically calculate the average fraction of frozen particles in rectangular systems of arbitrary dimensions for the Kob-Andersen and Fredrickson-Andersen kinetically-constrained models. We find the aspect ratio of the rectangle's length to width, which distinguishes short, square-like rectangles from long, tunnel-like rectangles, and show how changing it can effect the jamming transition. We find how the critical vacancy density converges to zero in infinite systems for different aspect ratios: for long and wide channels it decreases algebraically $v_{c}\sim W^{-1/2}$ with the system's width $W$, while in square systems it decreases logarithmically $v_{c}\sim1/\ln L$ with length $L$. Although derived for asymptotically wide rectangles, our analytical results agree with numerical data for systems as small as $W\approx10$.
\end{abstract}


\maketitle

\section{Introduction}

Increasing the density of particles in granular matter causes them to undergo a transition from an unjammed state, where the particles can move relatively freely, to a jammed state, where almost none of the particles can move~\cite{jamming}. Systems of interest in nature and in industrial applications typically have complicated geometries which strongly affect jamming in them~\cite{industry,industry2,nature,nature2}, and it is thus important to understand how does confinement influence the jamming of granular matter. Here we investigate the effects of confinement on the jamming transition, and in particular test how does the shape of containers determine how they jam. Most theoretical work so far was done on square systems \cite{durian,haxton,ohern,ningxu,lerner,barrat}.

There are numerous laboratory experiments that deal with non-square systems \cite{exp1,exp2,exp3,exp4}. For example, Daniels and Behringer conducted an experiment on polypropylene spheres in an annulus \cite{behringer1}, which is large enough to be considered a rectangle with infinite length and finite width. A different experiment by Bi et al. \cite{behringer2} consists of shearing a square system such that it becomes a rectangle with the same area and particle density as the original square. In this paper we study the effects of confinement on jamming by studying these phenomena in kinetically-constrained models in rectangular domains.

The essence of jamming is captured by the various kinetically-constrained models \cite{review,review2,kronig,fieldings,toninelli,sellitto,knights,DFOT,sellitto2,spiral,jeng,shokef,elmatad,driving}. For such simple models, and for other related models \cite{kipnis,derrida2,derrida,bouchaud,asep,monthus,nemeth,scheidler,varnik,srebro,zrp,lang}, it is useful to have exact solutions.

\begin{figure}[hb]
\includegraphics[width=160pt]{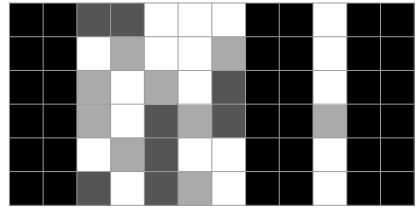}
\caption{
The difference between frozen and unfrozen particles in the Kob-Andersen model: White square are vacancies. Light-gray particles can move in this initial configuration. Dark-gray particles cannot move now, but after some other particle(s) move, they too are mobile. Black particles are permanently frozen and will never move.
}
\label{gridex}
\end{figure}

\subsection{Kinetically-Constrained Models}
Two types of kinetically-constrained models which describe granular and glassy materials are spin-facilitated models and lattice-gas models \cite{review,review2}. In both types of models, the system is represented by a grid, such that each site on the grid can have one of two values, $1$ or $0$. 

In the \textit{lattice-gas models}, a site with a value of $1$ represents a single particle and a site with a value of $0$ represents a vacant region. In each time step, one particle and one of the possible directions are chosen randomly with equal probabilities. The chosen particle attempts to move in that direction, and if the kinetic constraint allows the move, the particle moves to the neighboring site in the chosen direction. For a given initial configuration, some of the particles can move from the start, and some can move only after (many) other particles have moved and cleared the way for them. There may also be particles that will never move, no matter how the other particles in the system move. Those that will never move are called \textit{permanently frozen}, and those that can move eventually are called unfrozen. See Fig. \ref{gridex} for an example. 

In the case of \textit{spin-facilitated models}, a site with a value of $0$ represents a region of low density and high mobility, and a site with a value of $1$ represents a region of high density and low mobility. Note that this notation is different than the common notation ($0$ for a region of high density and $1$ for a region of low density), but we use this definition in order to deal with both lattice-gas and spin-facilitated models simultaneously. In each time step of the dynamics, one of the sites is chosen randomly, and changes its value at a temperature-dependent probability if it has enough neighbors with low density (i.e., a value of $0$), with the exact geometric criteria depending on the specific model. In the limit of zero temperature, the only allowed changes are from high density to low density, i.e. from a value of $1$ to $0$. For a given initial configuration, there is a possibility that even after an infinite number of time steps, some sites will still have a value of $1$. These permanently frozen sites represent the backbone of the system which will never change. 

In order to find the fraction of permanently frozen particles, one can use the bootstrap method, which iteratively removes mobile particles, until none of the remaining particles can move. Again, we have a backbone of sites which will never change. This algorithm is obviously valid for spin-facilitated models, but also for lattice gas models, since the criteria for the mobility of particles is local, and removing a mobile particle is effectively the same as moving it far enough from its neighbors. Since the algorithm for finding the backbone of both types of models is similar (but not identical), we will use the same language to describe both models, and choose the language of lattice-gas models. This means, for example, that whenever we speak of ``vacancies'' it should be interpreted as ``sites with value 0'' or ``low-density regions'' in the context of spin-facilitated models. For brevity, we will also use the term frozen particles interchangeably with permanently frozen particles.

We consider a two-dimensional rectangle, represented by a square lattice, such that each site either contains one particle or is vacant. The rectangle has $L$ sites in the horizontal direction, and $W$ sites in the vertical direction, such that $L\geq W$. In our numerical simulations we used hard-wall and periodic boundary conditions in both directions, but most of our analytical approximation ignores the boundary conditions. For rectangles of infinite length, hard-wall boundary conditions simulate particles inside a two dimensional channel, and periodic boundaries simulate particles on the surface of a cylindrical tube. 

For the lattice-gas model we use the Kob-Andersen (KA) model \cite{kamodel}, such that a particle can move if it has at least two neighboring vacancies before and after the move. For the spin-facilitated model we use the Fredrickson-Andersen (FA) model \cite{famodel}, such that a site can change its state if it has at least two neighboring sites with a value of $0$. We could have chosen a different number of neighbors needed for movement, but on the square lattice the only interesting effects occur at two neighbors. If only one neighbor is needed for movement then all the particles are movable as long as there is at least one vacancy in the FA model or two adjacent vacancies in the KA model. If three neighbors are needed for movement then any closed loop is frozen, even a $2\times2$ block, which means that almost all the particles in the system are frozen. Mathematically speaking, the KA model and the FA model are very similar to each other. Also, a mobile particle in the KA model is necessarily mobile in the FA model, and a frozen particle in the FA model is necessarily frozen in the KA model, thus the fraction of frozen particles in the KA model is larger than (or at least equal to) the fraction of frozen particles in the FA model. 

Toninelli, Biroli and Fisher showed \cite{toninelli} that for an infinite system in the KA model, none of the particles are permanently frozen as long as the lattice is not completely full with particles, which automatically means that this is also the case in the FA model. In this paper we study how many particles, on average, are permanently frozen for a given particle density, $\rho$, and given rectangle dimensions $W\times L$. 

\subsection{Finite Size Effects}

Numerical simulations done on square systems \cite{adler} $(L=W$, in our notations$)$ showed that the fraction of permanently frozen particles, $n_{PF}$ , rises rapidly from $0$ to $1$ at a certain critical density, $\rho_{c}$, which increases with system size. Holroyd~\cite{bp} theoretically analyzed jamming in this context using the notion of \textit{critical droplets}, which are small unjammed regions which facilitate movement throughout the system. He showed that for very large squares in the FA model the relation between the critical density and the system size is
\begin{align}
&\rho^{square}_{c}=1-\frac{\lambda}{\ln L} ,\nonumber\\
&\lambda=\frac{\pi^{2}}{18}\approx0.54 . \label{lambdadef}
\end{align}
Toninelli et. al. showed \cite{toninelli} that this value of $\lambda$ is also true for squares in the KA model.
However, this result is only true for asymptotically large $L$.

We can define an effective $\lambda$
\begin{align}
\lambda^{squares}_{eff}(L)=\left[1-\rho_{c}(L)\right]\ln L ,
\end{align}
which converges to $\lambda$ as the system size increases. For systems of size $L\approx10^{2}\sim10^{5}$ it was found \cite{aharony} that $\lambda^{squares}_{eff}\approx0.25$ for all simulated sizes. This contradiction was resolved by Holroyd's proof \cite{holroydslow} that the convergence of $\lambda^{squares}_{eff}$ to $\lambda$ is very slow and may be apparent only at systems of size $L\approx10^{20}$, beyond the capabilities of modern computers, and beyond the range of physical realization ($L=10^{20}$ implies a system of $L^{2}=10^{40}$ particles). 

Holroyd's analysis considered only the size of the system, and not its shape. For long rectangular domains, this method may not be used. Instead, we find that rectangular systems may be divided into independent \textit{sections}, in the sense that jamming in one section does not depend on the internal configuration within its neighboring sections. Within each section, Holroyd's notion of critical droplets may be used.

\subsection{Outline}

In this paper we show how not only the size of a system influences jamming in it, but also its shape. We demonstrate this, first by considering the limit of very large systems. When considering a square system of size $L \times L$, and taking the limit $L \rightarrow \infty$, one finds that the critical vacancy density, $v_{c}=1-\rho_{c}$, scales as $v_c \sim 1/\ln{L}$. We find that when the system's width $W$ is fixed and the length is taken to infinity $L=\infty$, the critical density scales as $v_c \sim 1/\sqrt{W}$ when $W\rightarrow\infty$ .

The second scenario we consider is of a system of fixed area, for which we change the aspect ratio between the width and the length of the system. We find that stretching the system causes it to jam, and relate this result to recent experiments of sheared granular matter.

The paper is organized as follows. In Section \ref{secdiv} of this paper we derive an approximate analytical expression for the fraction of frozen particles, $n_{PF}$, for a rectangular system of arbitrary dimensions $W\times L$. In Section \ref{seclarge} we deal with large systems, $W,L\gg1$, and use our approximation to find the critical density at which the system goes from jammed to unjammed and the width of this transition. We find that the system can be considered infinite if its length is longer than the average section length, or equivalently $\ln L\gg\sqrt{4\lambda W}$. In Section \ref{secsmall} we deal with narrow systems (small $W$), and improve the approximation derived in Section \ref{secdiv}. We even derive an exact result for the case of very narrow systems ($L=\infty$ and $W=1,2$). In Section \ref{secinter} we investigate the internal structures in the system. The Appendices contain the derivation of the lengthy expressions used in the analytical approximation.

\section{Critical Droplets and Division into Sections}
\label{secdiv}
\subsection{Critical Droplets}

Holroyd showed that in a large enough square there is a probability of approximately $e^{-2\lambda/v}$ that a particle is part of a critical droplet, where $\lambda$ is given in Eq. (\ref{lambdadef}b). Hence, the total number of critical droplets in a rectangle of size $L\times W$ is $WLe^{-2\lambda/v}$, where $v=1-\rho$ is the vacancy density. The expression for the critical density in square systems, Eq. (\ref{lambdadef}a), is derived by setting $W=L$. Such a critical droplet can cause the entire system to be unfrozen, thus the critical vacancy density is when the number of critical droplets is finite, since below (above) that critical density the number of critical droplets is very small (large) when $L$ is taken to infinity. 

The fraction of frozen particles in the FA and KA models is obviously different, due to the different kinetic constraints, but that difference is small. The reason that the densities of frozen particles for both the FA and the KA models are almost the same can be seen from the construction of Holroyd's proof. Holroyd considered small critical droplets, which are unfrozen, and checked how they can be expanded to ``unfreeze'' the entire system. The criterion for the expansion of these droplets is the same in both models, and the only difference is in what constitutes a small critical droplet. For example, the structure $\begin{array}{cc}1&0\\0&1\end{array}$ is unfrozen in the FA model but frozen in the KA model. For large enough systems, and evidently for small ones too, the effect of this difference is negligible.

However, when $L\rightarrow\infty$ and $W$ remains constant we cannot simply set $1=WLe^{-2\lambda/v_{c}}$ to find the critical density, since the solution to this equation is $v_{c}=0$ for all $W$. Our resolution of this problem is obtained by dividing the long rectangle into finite sections, implementing the idea of critical droplets in each section, and finally averaging over all sections. Another approach, which yields the same results, is solving the equation $1=W\left\langle l\right\rangle e^{-2\lambda/v_{c}}$, where $\left\langle l\right\rangle$ is the average section length (see below).

\subsection{Division into Sections}

\begin{figure}
\includegraphics[width=200pt]{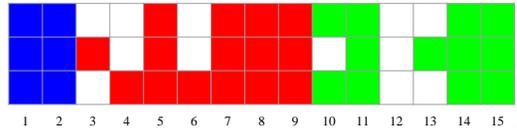}
\caption{(color online) 
Division into sections, represented by different colors: Columns $1-2$, $3-9$, and $10-15$. The strips are columns $1-2$, $7-9$, and $14-15$.
}
\label{gridex2}
\end{figure}

A rectangular system may be divided into sections by noting that if there are two or more successive columns which are completely full, then all the particles in them are permanently frozen in both the KA and FA models and with either hard-wall or periodic boundary conditions. We call a pattern of $m$ successive full columns a \textit{strip} of size $m$, where $m\geq2$. A single full column is not called a strip. These strips divide the rectangle into finite \textit{sections}, such that the leftmost column on each section is the first not-full column after a strip, and the rightmost column is the final column of the next strip. Each section contains only one strip. For example, see Fig. \ref{gridex2}. 

Using this division, we note that the particles within one section are independent from the particles in the other sections, i.e., the state of the particle (whether it is frozen or not) depends only on the structure within its section, and not on the configuration of neighboring sections.

Each finite section of length $l+m$ ending with a strip of size $m$ and with $n+mW$ occupied sites, has many configurations for the $n$ particles in the $l$ columns not occupied by the strip. We will denote these configurations with an index $s$. Since the probability of having a strip of length $m$ is independent of the probability for a certain configuration in the rest of the strip, the probability of such a configuration occurring is
\begin{align}
P(n,l,m,s)=\rho^{mW}Q(n,l,s), 
\end{align}
where $\rho^{mW}$ is the probability of having a strip of length $m$ containing $mW$ particles, and $Q(n,l,s)$ is the relative probability of configuration $s$ in the region with $l$ columns and $n$ occupied sites between the strips, such that there are no two adjacent full columns. The reason we exclude the possibility of two adjacent full columns is to count each type of section only once, since two (or more) adjacent full columns divide the section into smaller sections. The average fraction of frozen particles, $n_{PF}$, is the number of frozen particles divided by the number of particles,
\begin{align}
n_{PF}=\frac{\sum_{n,l,m,s}P(n,l,m,s)N(n,l,m,s)}{\sum_{n,l,m,s}P(n,l,m,s)\left[mW+n\right]}, \label{npfmain}
\end{align}
where $N(n,l,m,s)=N(n,l,s)+mW$ is the total number of permanently frozen particles in the section, with $N(n,l,s)$ being the number of frozen particles in the $l$ columns not occupied by the strip. The sum over $l$ and $m$ is such that $l+m\leq L$, and $m$ is greater or equal to $2$, except in the following special cases: no strip in the entire rectangle ($l=L,m=0$), and the entire section is full ($l=0,m=L$). Since the probability $P$ appears both in the nominator and the denominator in Eq. (\ref{npfmain}) we need not worry about its normalization or the normalization of $Q$. However, we find that in the limit of infinite length the probability is normalized such that $\sum P=1$.

\subsection{Our Approximation for Rectangular Systems}

\begin{figure}
\includegraphics[width=\columnwidth]{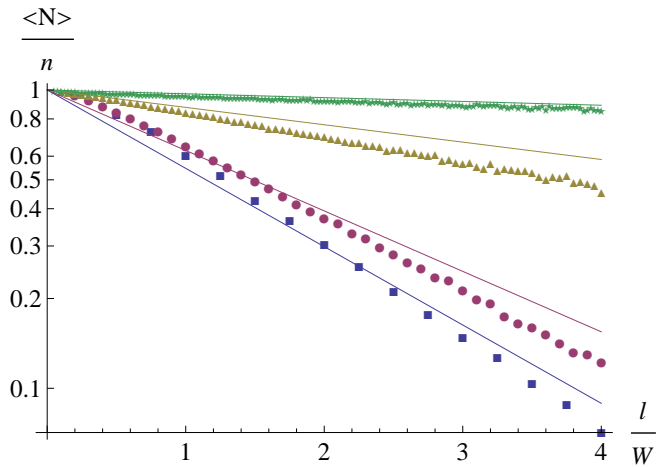}
\caption{(color online) 
Average fraction $\frac{\left\langle N\right\rangle}{n}$ of frozen particles in sections of length $l$ in the KA model with hard-wall boundary conditions for different widths and densities: $W=4$, $\rho=0.8$ (blue squares), $W=10$, $\rho=0.89$ (purple circles), $W=20$, $\rho=0.93$ (yellow triangles), $W=40$, $\rho=0.95$ (green stars). Continuous lines are approximations (\ref{krellam}), with $\lambda$ taken from simulations of long rectangles ($L=200W$).
}
\label{fakasec}
\end{figure}

In our case, we assume that the probability that a particle in a section of length $l+1$ is frozen is the probability that it is frozen in a section of length $l$ times the probability that the added column 
does not contain critical droplets, 
\begin{align}
\left\langle N(l+1)\right\rangle=\left\langle N(l)\right\rangle\left(1-e^{-2\lambda/v}\right)^{W} ,
\end{align}
where $\left\langle..\right\rangle$ is the average over all configurations. The solution to this recursion relation is
\begin{align}
&\left\langle N(n,l)\right\rangle=ne^{-kl} ,
&k=-W\ln\left(1-e^{-2\lambda/v}\right) .\label{krellam}
\end{align}
This leads to very good agreement with results of numerical simulations, as shown in Fig. \ref{fakasec}. 

Using this assumption in Eq. (\ref{npfmain}) yields
\begin{align}
n_{PF}=\frac{\sum_{n,l,m}\rho^{mW}Q(n,l)\left[mW+ne^{-kl}\right]}{\sum_{n,l,m}\rho^{mW}Q(n,l)\left[mW + n\right]} , \label{npfmain0}
\end{align}
where $Q(n,l)=\sum_{s}Q(n,l,s)$. The evaluation of these sums in closed form is given in Appendix A.
We note here that the denominator and the first part of the nominator in Eq. (\ref{npfmain0}) do not depend on our approximation relating the fraction of frozen particles with the number of critical droplets, and thus are not approximations but exact results. We further note that the ratio between the first part of the nominator and the denominator is the density of particles which are in the strips in both 
models and with both boundary condition. Hence, we define the density of particles which are in the strips as
\begin{align}
n_{strip}=\frac{\sum_{n,l,m}\rho^{mW}Q(n,l)mW}{\sum_{n,l,m}\rho^{mW}Q(n,l)\left[mW + n\right]}, 
\end{align}
which in the limit of $L\rightarrow\infty$ converges to
\begin{align}
n_{strip}(L\rightarrow\infty)=\rho^{2W-1}\left(2-\rho^{W}\right) .
\end{align}
We find that even for $W$ as small as $10$ and for all $L\geq W$, the density of particles in the strips is very low, except in the region very near $\rho=1$, where the fraction of frozen particles, $n_{PF}$, is almost unity. This means that for wide systems, the strips hardly contribute to the total fraction of frozen particles near the critical density and below it. The only role the strips play in this regime is dividing the system into sections, which are very long since there are few strips.
Using the results in Eqs. (\ref{denfinal}) and (\ref{nom1final}), the fraction of frozen particles at $L=\infty$ can be written as
\begin{align}
n_{PF}(L=\infty)=\rho^{2W-1}\left(2-\rho^{W}\right)+\frac{\rho^{4W-1}}{W}N_{PF}(\rho,W) , \label{npfex}
\end{align}
where
\begin{align}
N_{PF}(\rho,W)=\sum_{l,n,s}Q(l,n,s)N(l,n,s) .
\end{align}

\begin{figure}
\includegraphics[width=\columnwidth]{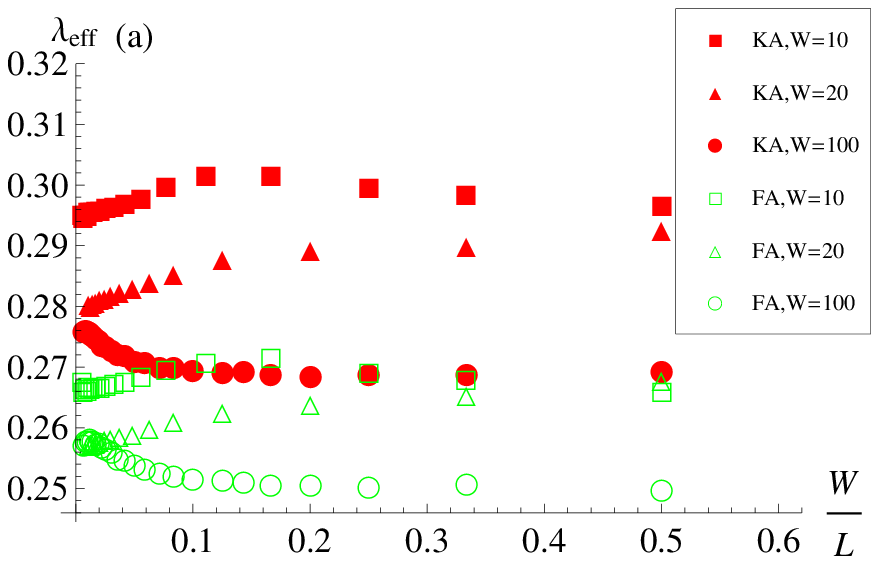}
\includegraphics[width=\columnwidth]{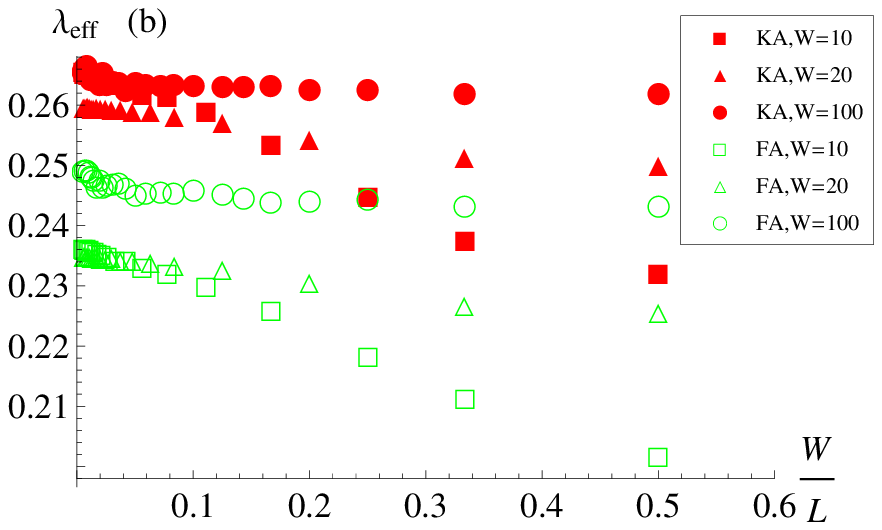}
\caption{(color online) The effective $\lambda$ as a function of $W/L$ for hard-wall (a) and periodic (b) boundary conditions. In this range, $\lambda_{eff}$ does not change much with the length $L$, and converges as $L\rightarrow\infty$ to a finite value for each width, $W$. For each system size, $\lambda_{eff}$ for the KA model is higher than in the FA model, and it is higher with hard-wall boundary conditions than with periodic boundary conditions, which means that $n^{KA}_{PF}>n^{FA}_{PF}$ and $n^{hard-wall}_{PF}>n^{periodic}_{PF}$, as expected. At $L\rightarrow\infty$, $\lambda$ is almost the same for $W=20$ and $W=100$ but different for $W=10$.}
\label{lameff}
\end{figure}

\begin{figure}
\includegraphics[width=\columnwidth]{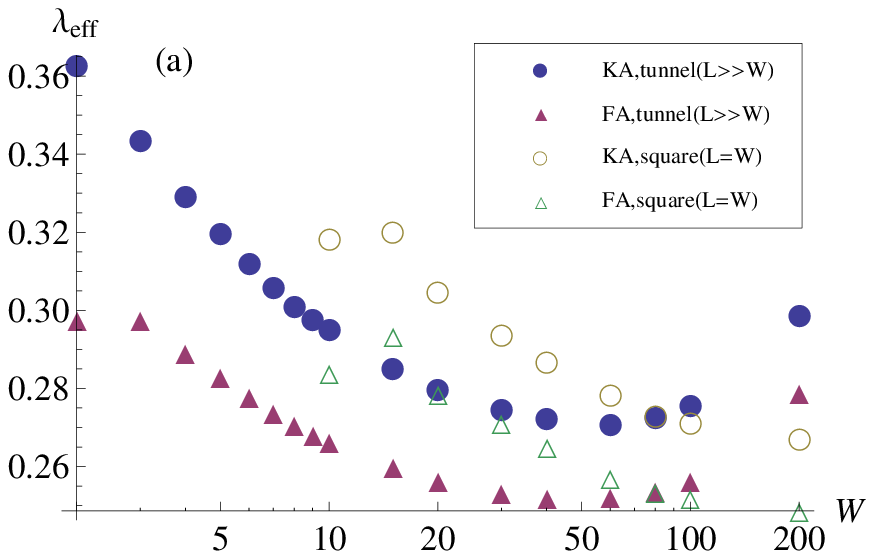}
\includegraphics[width=\columnwidth]{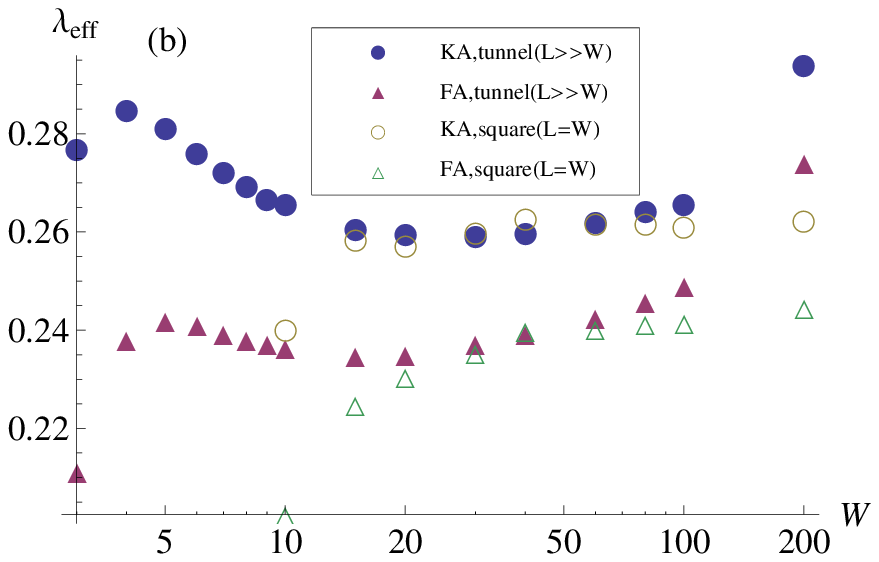}
\caption{(color online) The effective $\lambda$ as a function of the system's width $W$ for hard-wall 
(a) and periodic boundary conditions (b). The plots show the value of $\lambda_{eff}$ at $L=200W$ (for $W\leq100$) and $L=5000W$ (for $W=200$), which is large enough to be considered infinite in this range of widths, and at $L=W$. For $L\gg W$ the effective $\lambda$ has a minimum at $W\approx50$, but there is no drastic change in its value in the range $20\leq W\leq100$. Even at $W=200$ the relative difference from the minimum is $0.08$. 
$\lambda_{eff}$ for long tunnels ($L\gg W$) and squares ($L=W$) is almost the same at $60\leq W\leq100$. We expect that at $W\rightarrow\infty$, 
$\lambda$ will converge to 
$\pi^{2}/18\approx0.54$.}
\label{lameffinf}
\end{figure}

As previously shown for square systems, the value of $\lambda_{eff}$ depends on the system's size. We define the effective $\lambda$, $\lambda_{eff}(W,L)$, as the $\lambda$ for which the analytical approximation yields $n_{PF}(\rho_{c})=1/2$, with $\rho_{c}$ obtained by the numerical simulations. Previous simulations \cite{aharony} showed that for square systems in the FA model $\lambda_{eff}$ does not change much in the region 
$L\approx10^{2}-10^{5}$. 
Figure \ref{lameff} shows that for constant $W$, the value of $\lambda_{eff}(W,L)$ converges to a finite value $\lambda_{eff}(W,\infty)$, which is rather close to $\lambda_{eff}(W,W)$ at large $W$. See also Fig. \ref{lameffinf}. In the range of large $W$ and $L$ we see that the value of $\lambda$ depends mostly on the model, and not that much on the system's size or shape (square or rectangle). Figure \ref{lameffinf} shows the value of $\lambda$ for very long systems. We see from it that for squares it appears that $\lambda$ decreases with the width or converges to some value, but for long tunnels it is clear that the value of $\lambda$ increases with the width. Unless noted otherwise, in the rest of this paper we use the value of $\lambda$ taken from the simulations done on large squares.

\begin{figure}
\includegraphics[width=120pt]{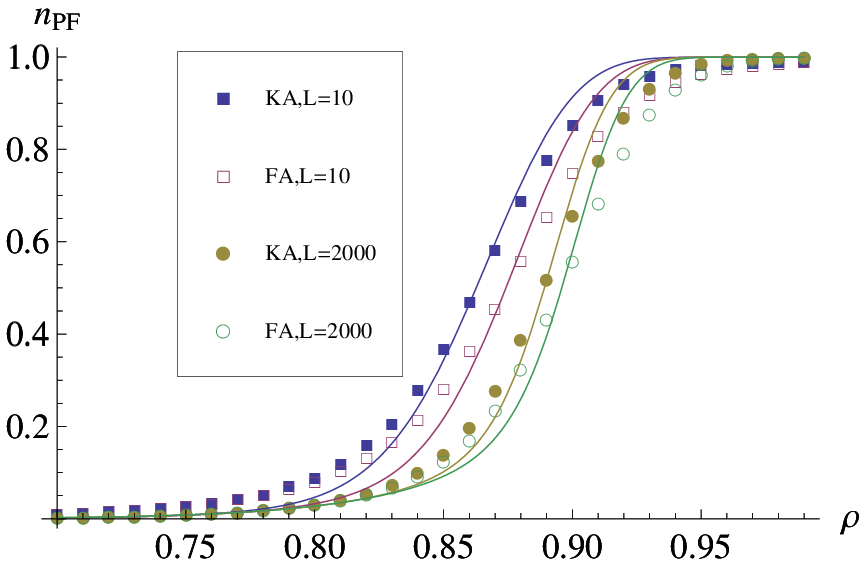}
\includegraphics[width=120pt]{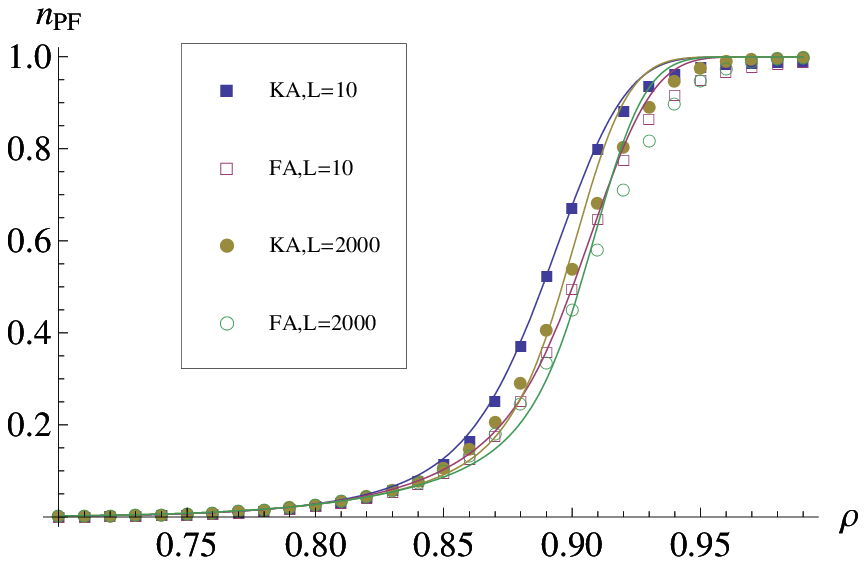}
\includegraphics[width=120pt]{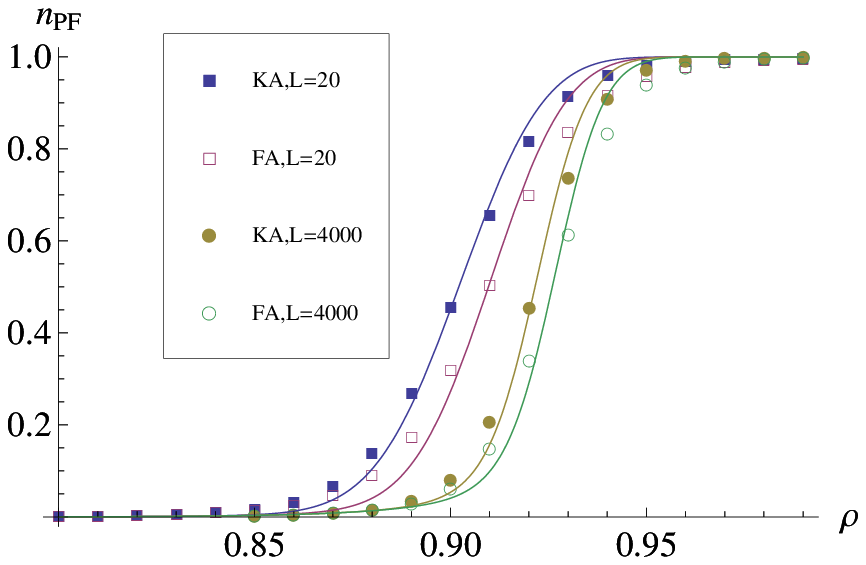}
\includegraphics[width=120pt]{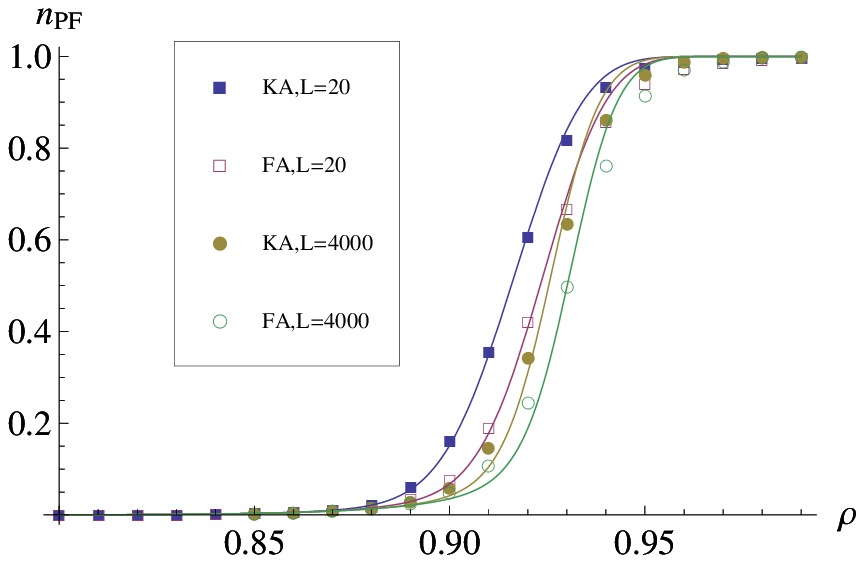}
\includegraphics[width=120pt]{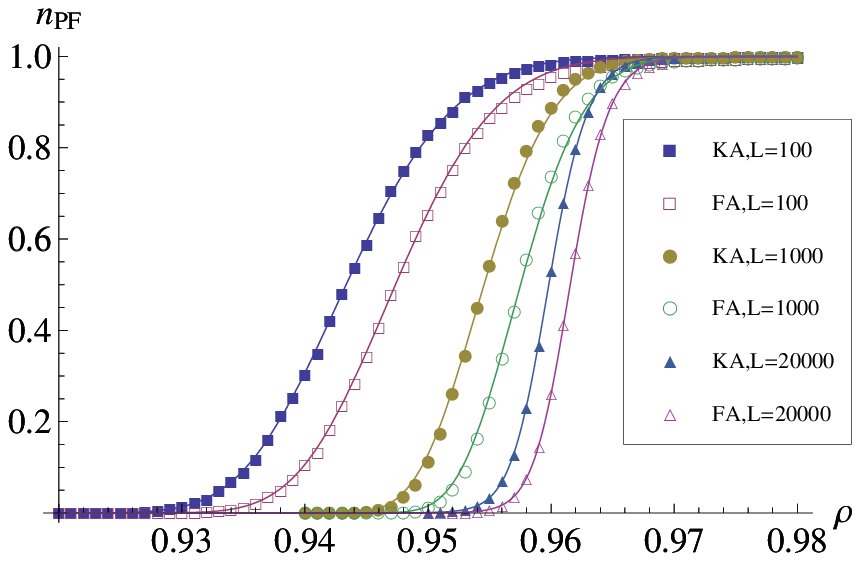}
\includegraphics[width=120pt]{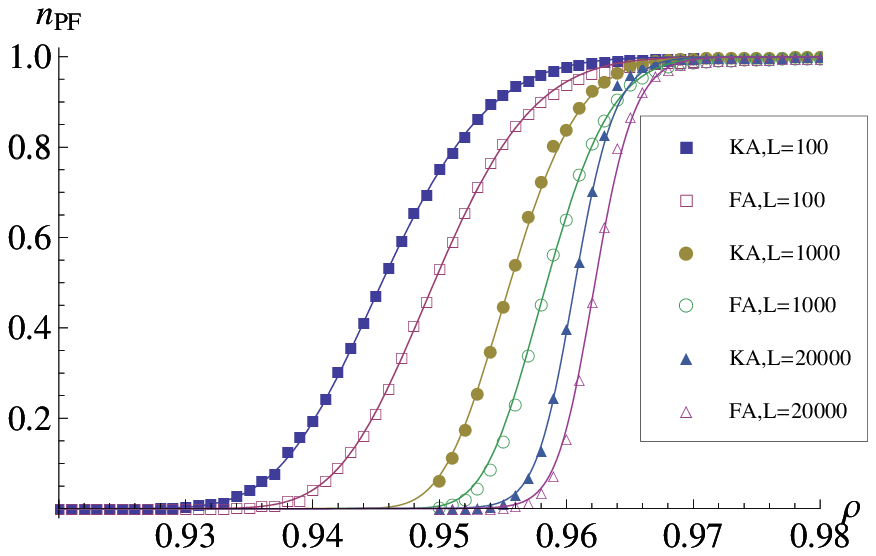}
\caption{(color online) 
Fraction of frozen particles $n_{PF}$ vs. density $\rho$ for hard-wall (left) and periodic (right) boundary conditions, and for different widths: $W=10$ (top), $W=20$ (middle), $W=100$ (bottom). Symbols are results of numerical simulations and continuous line is analytical approximation with different values of $\lambda_{eff}$. As the system's size increases, the approximation becomes better and the value of $\lambda_{eff}$ becomes dependent mostly on the model, and not on the system's size. $\lambda$ was set to $\lambda_{eff}(W,L)$, as shown in Fig. \ref{lameff}.
}
\label{npfw}
\end{figure}

Figure \ref{npfw} compares between the fraction of frozen particles, $n_{PF}$, obtained by the numerical simulations and the analytical approximation. From it we see that the approximation is roughly acceptable at $W=10$, quite good even at $W=20$, and has an excellent agreement with the numerical results at $W=100$, even for the hard-wall boundary conditions. Also, we note that the approximation is better for the periodic boundary conditions, and that the width of the transition from an unjammed state, where almost all of the particles are unfrozen, to a jammed state, where almost all the particles are frozen, is narrower with the periodic boundary conditions.

\section{Large Systems, $W,L\gg1$}
\label{seclarge}
In large systems, the transition from jammed ($n_{PF}\approx1$) to unjammed ($n_{PF}\approx0$) occurs in a very narrow region of densities. In what follows we find the critical density, $\rho_{c}$, at which this transition occurs and the width of the transition, $\Delta\rho$. We also show how the critical density depends on the shape of the system and not only on its size, by considering a system of fixed area and changing the aspect ratio $W/L$. Note that for finite-sized systems (and even when $L$ is infinite but $W$ is finite), there is no singularity in any physical quantity. Nonetheless we use the term critical density since permanently frozen particles exist due to the same considerations that govern jamming in the thermodynamic limit, where one may discuss the notion of critical phenomena \cite{durian,haxton,olsson}.

\subsection{Critical Density}
\subsubsection{Critical Density from Fraction of Frozen Particles}
We first note that as the system grows larger, the critical density grows as well and nears $1$. Therefore, we can use the known critical density for squares of size $W\times W$, Eq. (\ref{lambdadef}), to find a lower bound on the critical density in rectangles of size $W\times L$ ($W\leq L$)
\begin{align}
v_{c}\leq\frac{\lambda}{\ln W} .\label{boundvc}
\end{align}

From Eq. (\ref{boundvc}) we find that
\begin{align}
&k=-W\ln\left(1-e^{-2\lambda/v_{c}}\right)\leq We^{-2\ln W}=W^{-1}\ll1 .
\end{align}
Since $W$ is very large, this means that the exponent in the logarithm is very small, and thus
\begin{align}
k\approx We^{-2\lambda/v} .
\end{align}
We also note that the critical density is very close to $1$ but still $\left(\rho_{c}\right)^{W}\ll1$. Therefore, close to the critical density, we can use the results from Appendix \ref{apnpf} and approximate $n_{PF}$ by
\begin{align}
&n_{PF}\approx\nonumber\\
&\frac{1+\exp\left[-L\rho^{2W}\left(k\rho^{-2W}+1\right)\right]\left[kL\left(k\rho^{-2W}+1\right)-1\right]}{\left(k\rho^{-2W}+1\right)^{2}
\left[1-\exp\left(-L\rho^{2W}\right)\right]
} .
\end{align}

In very short rectangles, such that $L\rho^{2W}\ll1$, we find that $n_{PF}$ is finite only if $k\rho^{-2W}\gg1$ and $kL$ is finite, and thus
\begin{align}
n_{PF}(L\rho^{2W}\ll1)\approx e^{-kL} .\label{npfshort}
\end{align}
Solving the equation $n_{PF}=1/2$ yields
\begin{align}
v_{c}\approx\frac{2\lambda}{\ln\left(WL\right)} ,\label{vcshort}
\end{align}
which retrieves the known result, Eq. (\ref{lambdadef}), for the case $W=L\gg1$.

In very long rectangles, such that $L\rho^{2W}\gg1$, we find that $n_{PF}$ is finite only if $k\rho^{-2W}$ is finite and thus
\begin{align}
n_{PF}(L\rho^{2W}\gg1)\approx\frac{1}{\left(k\rho^{-2W}+1\right)^{2}} .\label{npflong}
\end{align}
Solving the equation $n_{PF}=1/2$ yields
\begin{align}
v_{c}\approx\frac{\sqrt{16\lambda W+\ln^{2}\left(W\right)}-\ln\left(W\right)}{4W} .\label{vclong}
\end{align}

This means that the distinction between short and long rectangles is whether $L\left(\rho_{c}\right)^{2W}$ is greater or lesser than $1$. Equating Eqs. (\ref{vcshort}) and (\ref{vclong}), we find that the crossover from short rectangles to long rectangles occurs at
\begin{align}
\ln L_{c}=2Wv_{c}=\frac{\sqrt{16\lambda W+\ln^{2}\left(W\right)}-\ln W}{2} .\label{lcbig}
\end{align}

\subsubsection{Critical Density from Typical Section Length}

Another approach for finding the critical density is by considering only a typical section of length $\left\langle l\right\rangle$, where
\begin{align}
\left\langle l\right\rangle=\frac{\sum^{L}_{l=1}Q(l)l}{\sum^{L}_{l=1}Q(l)} 
\end{align}
is the average section length. In this case, similarly to what was done on square systems, we need to solve the equation
\begin{align}
1=W\left\langle l\right\rangle e^{-2\lambda/v_{c}} .\label{eqav}
\end{align}
Using the expressions in the Appendix, and assuming that $\rho^{W}\ll1$ and $L,W\gg1$, we find that
\begin{align}
\left\langle l\right\rangle\approx\rho^{-2W}
\left[1+\frac{L\rho^{2W}}{1-\exp\left(L\rho^{2W}\right)}\right]
.\label{lav}
\end{align}
In the limits of $L\rho^{2W}\gg1$ and $L\rho^{2W}\ll1$, solving Eq. (\ref{eqav}) yields the same results as in Eqs. (\ref{vcshort}) and (\ref{vclong}).

This means that when the rectangle's length is shorter than $\left\langle l\right\rangle$ it can be considered as consisting of a single section, and that for longer rectangles we can consider only sections of average length. For this reason short, square-like rectangles can be considered to contain only one section and the critical density in them depends as a first approximation only on the system area $WL$ and not on its shape. 

By considering terms of order $L\rho^{2W}=L/L_{c}$, we find that the correction to Eq. (\ref{vcshort}) is
\begin{align}
v_{c}\approx\frac{2\lambda}{\ln\left(WL\right)}+\frac{L}{L_{c}}\frac{\lambda}{\ln^{2}\left(WL\right)}\frac{2-\ln^{2}(2)-\ln(4)}{\ln^{3}(2)} .\label{vcshort2}
\end{align}
By keeping the ratio $L/L_{c}$ constant, we see from Eq. (\ref{vcshort2}) that the correction becomes less important at larger systems.

\subsubsection{Alternative Approach on Confinement}

\begin{figure}
\includegraphics[width=\columnwidth]{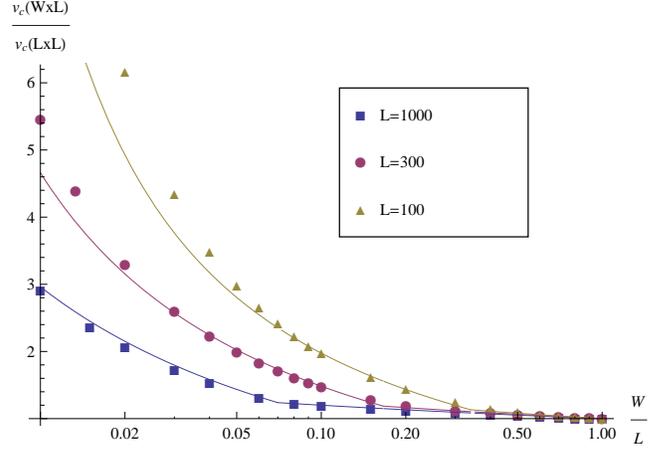}
\caption{(color online) 
Ratio between critical vacancy density in rectangles of dimension $W\times L$ and critical vacancy density in squares of dimension $L\times L$ in the KA model with hard-wall boundary conditions. Symbols are results of numerical simulations and continuous line is the approximations (\ref{vcshort}) and (\ref{vclong}). The ratio between the critical densities significantly differs from $1$ only for $W<W_{c}(L)$. $W_{c}/L$ decreases with $L$, and the ratio between the critical densities increases with $L$.
}
\label{vcrel}
\end{figure}

So far we looked at what happens when the width remains constant and the length increases, and found the crossover length $L_{c}$ between the two limiting cases described above. Another way to look at it is by starting from a square of size $L\times L$ and to generate confinement by narrowing its width. Namely, we fix $L$, and decrease $W$. In this case we find from Eq. (\ref{lcbig}) a crossover width
\begin{align}
W_{c}=\frac{\ln^{2}(L)}{4\lambda} ,\label{wc}
\end{align}
which may be interpreted as the width below which the system can be considered infinitely long. To find the effect of the confinement on the critical density, we calculate the ratio between the critical density in squares and in rectangles. For $W>W_{c}(L)$ or equivalently for $L<L_{c}(W)$, this ratio is
\begin{align}
\frac{v_{c}(W\times L)}{v_{c}(L\times L)}\approx\frac{2}{1+\frac{\ln(W)}{2\sqrt{\lambda W_{c}}}}\approx\frac{2}{1+\frac{2\ln\ln(L_{c})}{\ln L}} ,
\end{align}
and for $W<W_{c}(L)$,
or $L>L_{c}(W)$,
the ratio is
\begin{align}
\frac{v_{c}(W\times L)}{v_{c}(L\times L)}\approx2\sqrt{\frac{W_{c}}{W}}\approx\frac{2\ln(L)}{\ln(L_{c})} .
\end{align}
Figure \ref{vcrel} shows the ratio between the critical vacancy density in rectangles of size $W\times L$ and the critical vacancy density in squares of size $L\times L$. We see from it that the asymptotic values in Eqs. (\ref{vcshort}) and (\ref{vclong}) agree with the numerical results even at $W\approx W_{c}$. This means that systems really may be divided into long and short rectangles with a clear distinction between the two types.

\begin{figure}
\includegraphics[width=\columnwidth]{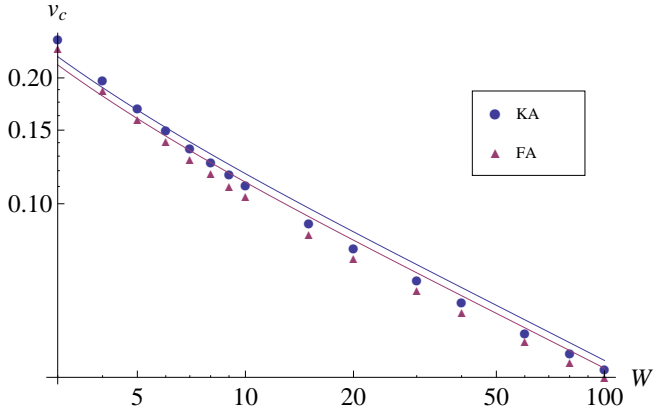}
\caption{(color online) The critical vacancy density, $v_{c}$, as a function of the width for very long systems ($L=200W\gg L_{c}(W)$) with hard-wall boundaries. The symbols are the results from the numerical simulations and the continuous lines are the approximation (Eq. (\ref{vclong})) with $\lambda=\lambda_{eff}(100,100)$.}
\label{vcinf}
\end{figure}

The critical density in Eq. (\ref{vclong}) is a very good approximation even for $W$ as small as $3$, as shown in Fig. \ref{vcinf}. This dependence of the critical density on the width of the system in long channels can be measured in experiments. The suggested value for the effective $\lambda$ is $0.257(FA),0.275(KA)$ for hard-wall boundaries and $0.249(FA),0.271(KA)$ for periodic boundaries, since this is its value for systems of infinite length and with a width of $W\approx20-100$.

\subsection{Varying Aspect Ratio}

Here we consider a system of fixed volume $V=WL$, and study how changing the aspect ratio between the width and the length, $r=W/L$, effects the jamming transition. The crossover aspect ratio, $r_{c}$, is the 
aspect
ratio which defines whether the system behaves as a square-like system or 
as
a long system. From Eq. (\ref{wc}) we find that the crossover aspect ratio satisfies the equation
\begin{align}
16\lambda\sqrt{Vr_{c}}=\ln^{2}\left(r_{c}/V\right) .\label{rc}
\end{align}
If the density is high enough that the system is jammed at $r=1$, then it is also jammed at any other $r<1$. If the system is not jammed at $r=1$, the density can still be the critical density at some aspect ratio smaller than $1$. This means that as the aspect ratio decreases, the system may undergo a jamming transition if the density is below the critical density at $r=1$ but above the critical density at $r\ll1$. 

\begin{figure}
\includegraphics[width=\columnwidth]{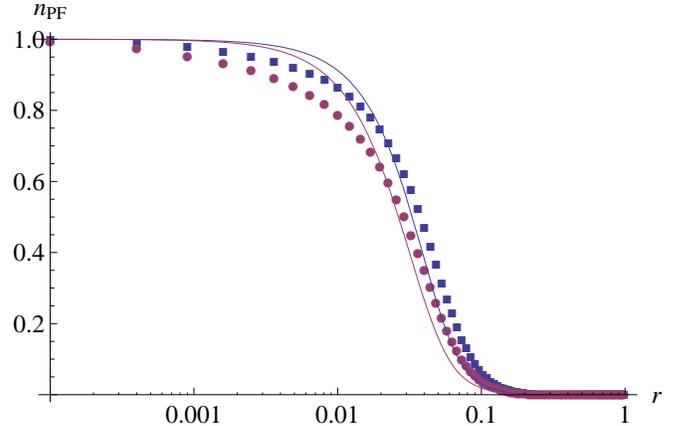}
\caption{(color online) 
Fraction of frozen particles vs aspect ratio $r=W/L$ at constant area $WL=10^{4}$ and particle density $\rho=0.92$ with hard-wall boundaries. Symbols are results of numerical simulations for KA (blue squares) and FA (purple circles) models. Continuous lines are analytical approximations with $\lambda_{eff}(100,100)\approx0.271$ (KA), 
$0.252$ (FA).
}
\label{v104}
\end{figure}

For example, we now show that our results may be related to recent experiments of Bi et al. \cite{behringer2}. In these experiments shear stress was applied on a two-dimensional system, such that its area and particle density remained constant, but the aspect ratio between its length and width changed. We 
consider a system of fixed area $WL=10^{4}$ and density $\rho=0.92$ with hard-wall boundaries. The density was chosen such that it is below $\rho_{c}$ for $r=1$ and above $\rho_{c}$ for $r\ll1$. By changing the aspect ratio, the fraction of frozen particles changes from almost $0$ at $r\geq0.1$ to almost $1$ at $r\leq0.01$ as seen in Fig. \ref{v104}. The crossover aspect ratio from Eq. (\ref{rc}) is $r_{c}=0.093$, very near the $r=0.1$ observed in the numerical results. This means that shearing the system causes it to become jammed, in agreement with the experimental results \cite{behringer2}. 

\subsection{Transition Width}

We define the width of the transition, $\Delta\rho$ as the difference between the densities for which $n_{PF}=\delta$ and $1-\delta$, where $\delta$ is an arbitrary number much smaller than $1$. In short rectangles
$[L\left(\rho_{c}\right)^{2W}\ll1]$, 
setting $n_{PF}$ in Eq. (\ref{npfshort}) equal to $\delta\ll1$ and to $1-\delta$ yields
\begin{align}
&\Delta\rho(short)=\rho_{u}-\rho_{l}\approx\nonumber\\
&\approx\frac{2\lambda}{\ln^{2}\left(WL\right)}\ln\left(\frac{\ln\delta^{-1}}{\delta}\right)\approx\frac{v^{2}_{c}\ln\delta^{-1}}{2\lambda} ,\label{dr}
\end{align}
where $\rho_{l,u}$ are the values of the density at the lower $(n_{PF}=\delta)$ and the upper $(n_{PF}=1-\delta)$ bounds.
For long rectangles 
$[L\left(\rho_{c}\right)^{2W}\gg1]$, 
setting $n_{PF}$ in Eq. (\ref{npflong}) equal to $\delta$ and to $1-\delta$ yields
\begin{align}
&\Delta\rho(long)=\rho_{u}-\rho_{l}\approx\frac{3\ln\delta^{-1}}{8W}\approx\frac{3v^{2}_{c}\ln\delta^{-1}}{8\lambda} ,
\end{align}
which slightly differs from Eq. (\ref{dr}) only in the numerical prefactor.

\begin{figure}
\includegraphics[width=\columnwidth]{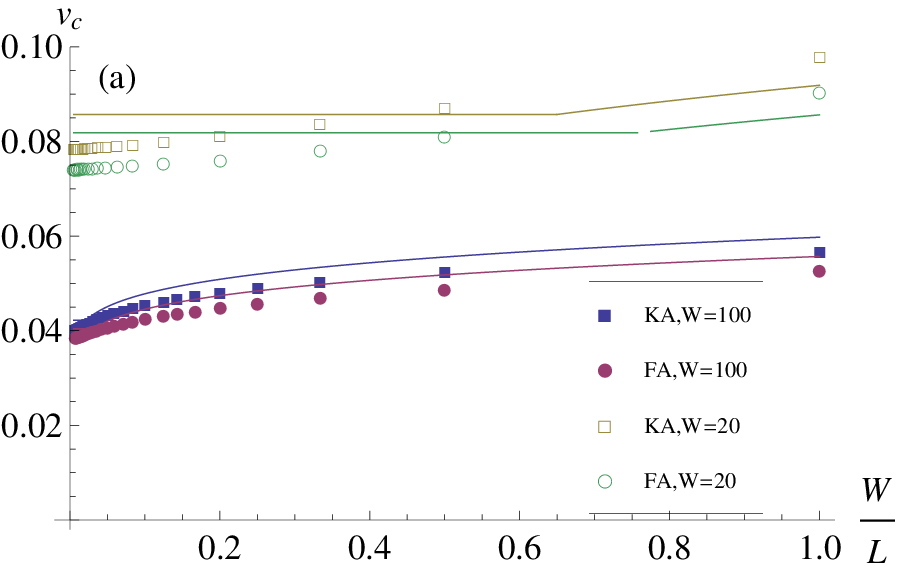}
\includegraphics[width=\columnwidth]{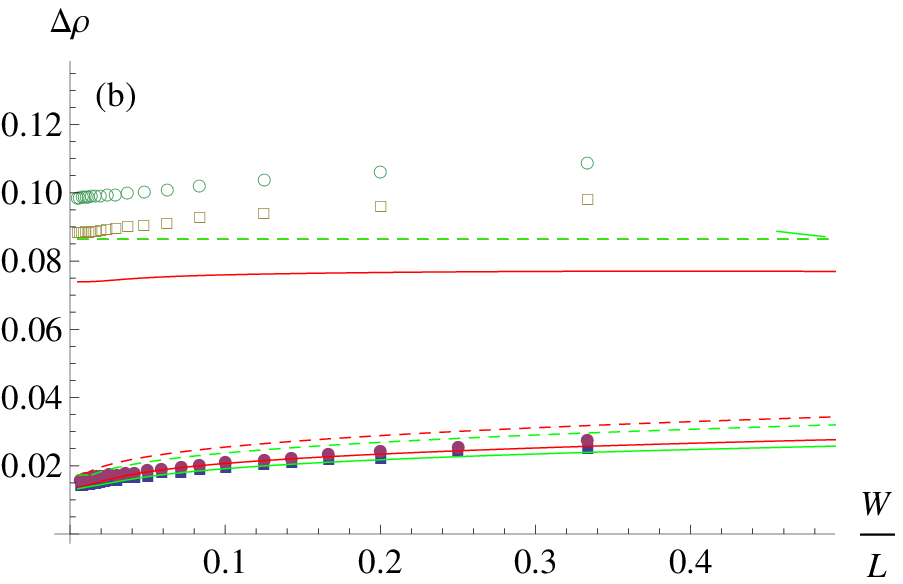}
\caption{(color online) 
Critical vacancy density (a) and transition width (b) with hard-wall boundaries. Symbols are the same in both panels. In panel (a), full lines are the approximations (\ref{vcshort},\ref{vclong}) for large $W$. In panel (b), full lines are the results from the full analytical expression and the dashed lines are the large-$W$ approximations.
}
\label{rcbig}
\end{figure}

Figure \ref{rcbig} shows the critical density and width of the transition for systems of width $W=100$. We see, for example for $W=100$, that the critical density and the transition width indeed converges at $L_{c}\approx5000(KA),3000(FA)$, in agreement with Eq. (\ref{lcbig}) which predicts $L_{c}=4606(KA),3152(FA)$. For the smaller width, $W=20$, there is also a convergence, but the fit is not as good as for $W=100$. The difference between the numerical results and the approximations for large $W$ is visible because $W=100$ is still not large enough for the asymptotic limit.

\section{Systems with Small Width}
\label{secsmall}
In systems of small width we can improve our approximation, and even have exact results. In the first two subsections below we find an exact result for $n_{PF}$ for $L=\infty$ and $W=1$ or $2$ in both the KA and FA models. In the third subsection we improve our approximation for systems of width $3\leq W\leq 6$ by finding the exact number of frozen particles in small sections.

\subsection{Fraction of Frozen Particles at $W=1$ and $L=\infty$}
In systems of width $W=1$ we note that the boundary conditions are irrelevant. With hard-wall boundaries, a particle is blocked from above and below by the boundaries, and with periodic boundaries it is blocked by itself.

\subsubsection{KA Model}
For systems with $W=1$, we denote by $f$ the number of occupied sites in the section and by $h$ the number of vacancies in the section. As there are no two adjacent occupied sites in the section, in order for a particle to be permanently frozen in the KA model, the entire section must be frozen and moreover it must be of the form $01010...01010$, i.e. $h=f+1$. The number of frozen particles in systems with $W=1$ is thus
\begin{align}
N^{KA}_{PF}(\rho,1)=\sum^{\infty}_{h=1}v^{h}\rho^{h-1}\left(h-1\right)=\frac{\rho v^{2}}{\left(1-\rho v\right)^{2}} ,
\end{align}
and thus, using Eq. (\ref{npfex}), the fraction of frozen particles is
\begin{align}
n^{KA}_{PF}(\rho,1)=2\rho-\rho^{2}+\frac{\rho^{4}v^{2}}{\left(1-\rho v\right)^{2}} .
\end{align}

\subsubsection{FA Model}
In the FA model, only the particles in the strips are frozen for $W=1$, and thus
\begin{align}
n^{FA}_{PF}(\rho,1)=2\rho-\rho^{2} .
\end{align}

\subsection{Fraction of Frozen Particles at $W=2$ and $L=\infty$}
For systems with $W=2$ the boundary conditions are important.

\subsubsection{KA Model, Hard-Wall Boundaries}
Consider the following two types of patterns: First, a pattern with zigzag diagonals of occupied sites, such that the other sites are either occupied or not, and second, a pattern with a full row of occupied sites, with the sites in the other row either occupied or not. In each of these two cases, the particles in the \textit{main part} (the full zigzag or full row) are frozen, and in the \textit{secondary part} they are either frozen (if the main part is a zigzag) or unfrozen (if the main part is a row). A section with frozen particles can be built by dividing it into subsections with their main part either a zigzag or a row. Each two of these subsections must be divided by a \textit{divider}, which consists of a full column and one particle in each of the adjacent columns, one at the top and one at the bottom. The following scheme shows this more clearly:
\begin{align}
\begin{array}{cccccccccccccccccc}
1&0&1&0&d&d&0&0&0&2&0&0&d&d&3&3&3&3\\
0&1&0&1&0&d&d&2&2&2&2&d&d&0&3&0&3&0
\end{array}
\end{align}
The sites marked with $d$ belong to a divider, and the sites marked with a number belong to one of the subsections. The first subsection is of the zigzag type. The second subsection is of the row type. Note that the site marked $2$ at the top row is not frozen. The third subsection can be of either type. We will consider it to be of a zigzag type, since all the particles in the secondary part are frozen. In order to simplify the following calculations, we will include the rightmost column of a divider in the subsection to the right of it. By denoting $d$ as the number of dividers, we note that the number of subsections is $d+1$. We account for the possibility of two adjacent dividers by considering subsections of length $0$. Also, since the left column in the divider is counted in it, we need to artificially add the rightmost column in the rightmost subsection, since it is not counted in the (non-existent) divider to the right of the last subsection. Another point to make is that the leftmost column in a subsection cannot be full. We denote each subsection by the number of vacancies, $h_{i}$, and the number of occupied sites, $f_{i}$, in the secondary part, such that the main part contains $h_{i}+f_{i}$ sites, and by its type, $t_{i}=z,r$ (zigzag or row). A particle in a section built in this way is unfrozen only if it is in the secondary part of a subsection of row type. We also need to make sure there are no two adjacent columns. Also, for each such section, there is a mirror configuration with the top and bottom rows switched, and so we can count the number of frozen particles in one such configuration (say, with the occupied site on the leftmost column in the top row) and multiply by $2$. 

The number of frozen particles in such a section is thus
\begin{widetext}
\begin{align}
N^{KA,hw}_{PF}(\rho,2)=2\sum^{\infty}_{d=0}\left(\rho^{3}v\right)^{d}\prod^{d+1}_{i=1}\sum_{t_{i}=z,r}\sum^{\infty}_{h_{i}=0}\sum^{h_{i}-\delta_{t_{i},r}}_{f_{i}=0}v^{h_{i}}\rho^{h_{i}+2f_{i}}\left(\begin{array}{c}h_{i}\\f_{i}\end{array}\right)\rho v\left[3d+\sum^{d+1}_{j=1}\left(h_{j}+f_{j}+f_{j}\delta_{t_{j},z}\right)+1\right] .\label{w2eq1}
\end{align}
The factor of $2$ at the beginning is for the top-bottom symmetry. In the sums, we go over each subsection and check how many vacancies and occupied sites there are in the secondary part, where we note that in a subsection of row type $f_{i}<h_{i}$ (otherwise we consider it a zigzag type). The factor of $\rho v$ before the square brackets is for the rightmost column. The sum in the square brackets requires more explanations. First we add the particles in the main part ($h_{i}+f_{i}$). Next, we say that an occupied site in the secondary part is frozen only if the subsection is of zigzag type. The $1$ at the end is for the particle in the rightmost column. 
The final result from evaluating the sums in Eq. (\ref{w2eq1}) is (see Appendix \ref{apka2})
\begin{align}
&N^{KA,hw}_{PF}(\rho,2)=\frac{2\rho v\left(1-\rho^{3}v\right)\left[\left(1-\rho^{3}v\right)\left(1+4\rho^{3}v\right)+\rho v\left(2-\rho v+6\rho^{3}v\right)\right]}{\left[\left(1-\rho^{3}v\right)\left(1-2\rho^{3}v\right)-\rho v\right]^{2}} ,\label{ka2end}
\end{align}
and the fraction of frozen particles is
\begin{align}
&n^{KA,hw}_{PF}(\rho,2)=\rho^{3}\left(2-\rho^{2}\right)+\frac{\rho^{7}}{2}N^{KA,hw}_{PF}(\rho,2) . \label{ka2hw}
\end{align}

\subsubsection{FA Model, Hard-Wall Boundaries}
In the FA model with hard-wall boundaries, a section can be at least partially frozen only if all of its subsections are of row type, and the frozen particles are only those in the dividers and in the main part. Thus, the number of frozen particles is (see Appendix \ref{apfa2})
\begin{align}
N^{FA,hw}_{PF}(\rho,2)=2\sum^{\infty}_{d=0}\left(\rho^{3}v\right)^{d}\prod^{d+1}_{i=1}\sum^{\infty}_{h_{i}=0}\sum^{h_{i}}_{f_{i}=0}v^{h_{i}}\rho^{h_{i}+2f_{i}}\left(\begin{array}{c}h_{i}\\f_{i}\end{array}\right)\rho v\left[3d+\sum^{d+1}_{j=1}\left(h_{j}+f_{j}\right)+1\right]=\frac{2\rho v\left(1+3\rho^{3}v\right)}{\left[1-\rho v\left(1+2\rho^{2}\right)\right]^{2}} . \label{fa2begin}
\end{align}
The fraction of frozen particles is thus
\begin{align}
n^{FA,hw}_{PF}(\rho,2)=\rho^{3}\left(2-\rho^{2}\right)+\frac{\rho^{8}v\left(1+3\rho^{3}v\right)}{\left[1-\rho v\left(1+2\rho^{2}\right)\right]^{2}} . \label{fa2hw}
\end{align}

\subsubsection{KA Model, Periodic Boundaries}
In the KA model with periodic boundaries, a section is frozen only if all of its subsections are of zigzag type, and thus the number of frozen particles is (see Appendix \ref{apkap2})
\begin{align}
N^{KA,per}_{PF}(\rho,2)=2\sum^{\infty}_{d=0}\left(\rho^{3}v\right)^{d}\prod^{d+1}_{i=1}\sum^{\infty}_{h_{i}=0}\sum^{h_{i}}_{f_{i}=0}v^{h_{i}}\rho^{h_{i}+2f_{i}}\left(\begin{array}{c}h_{i}\\f_{i}\end{array}\right)\rho v\left[3d+\sum^{d+1}_{j=1}\left(h_{j}+2f_{j}\right)+1\right]=\frac{2\rho v\left(1+4\rho^{3}v\right)}{\left[1-\rho v\left(1+2\rho^{2}\right)\right]^{2}} ,\label{ka2per}
\end{align}
and the fraction of frozen particles is
\begin{align}
n^{KA,per}_{PF}(\rho,2)=\rho^{3}\left(2-\rho^{2}\right)+\frac{\rho^{8}v\left(1+4\rho^{3}v\right)}{\left[1-\rho v\left(1+2\rho^{2}\right)\right]^{2}} . \label{ka2p}
\end{align}
\end{widetext}

The reason that Eqs. (\ref{fa2hw}) and (\ref{ka2p}) are very similar is that we count almost the same number of particles. The only addition to Eq. (\ref{ka2p}) from Eq. (\ref{fa2hw}) is the particles in the secondary part.

\subsubsection{FA Model, Periodic Boundaries}
In this case, the only frozen particles are in the strips, and thus
\begin{align}
&n^{FA,per}_{PF}(\rho,2)=\rho^{3}\left(2-\rho^{2}\right). \label{fa2p}
\end{align}

\begin{figure}
\includegraphics[width=\columnwidth]{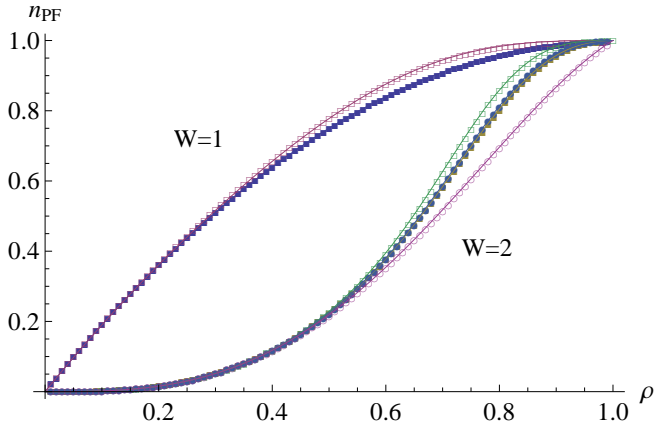}
\caption{(color online) 
Fraction of frozen particles vs. density for the FA (full symbols) and KA (empty symbols) models, for hard-wall (squares) and periodic (circles) boundary conditions, for $W=1,2$. Symbols are results of numerical simulations with $L=200W$, continuous line is analytical expression. For $W=2$ the results for KA with periodic boundaries and FA with hard-wall boundaries are almost the same.
}
\label{compw12}
\end{figure}

Figure \ref{compw12} shows how perfectly the expressions (\ref{ka2hw}), (\ref{fa2hw}), (\ref{ka2p}), and (\ref{fa2p}) fit the numerical results.

\subsection{Fraction of Frozen Particles at $3\leq W\leq6$}
\label{sec36}
In principle, we can find exact results also for systems with $W>2$, but as seen in the previous subsections it gets progressively more complicated with increasing $W$. However, we can improve the approximation calculated previously. For systems with small width ($3\leq W\leq6$) we can calculate exactly the average number of frozen particles in sections of length $l$, $N(l)$, by simply counting all possible configurations. As the number of possible configurations rises exponentially with the section's length and the system's width, we will consider this only for sections of size $Wl\leq40$, for which the number of configurations is $2^{40}\approx10^{12}$, a number which can be handled numerically. The frozen particles in the longer sections are neglected.

\begin{figure}
\includegraphics[width=120pt]{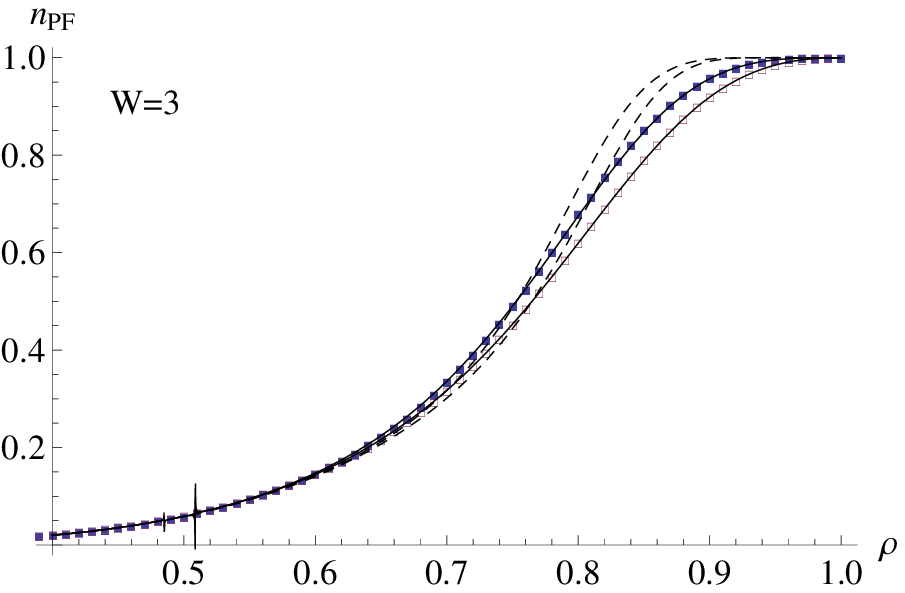}
\includegraphics[width=120pt]{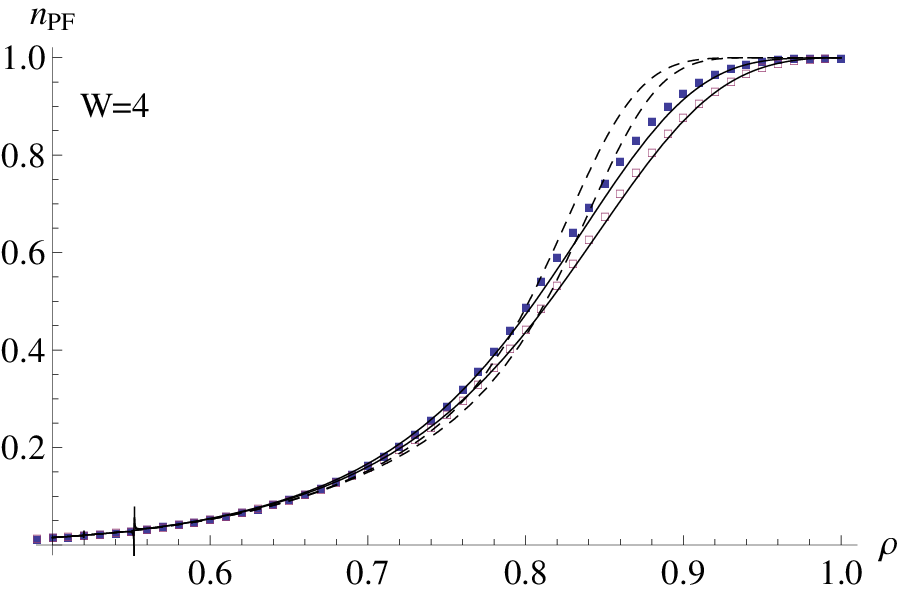}
\includegraphics[width=120pt]{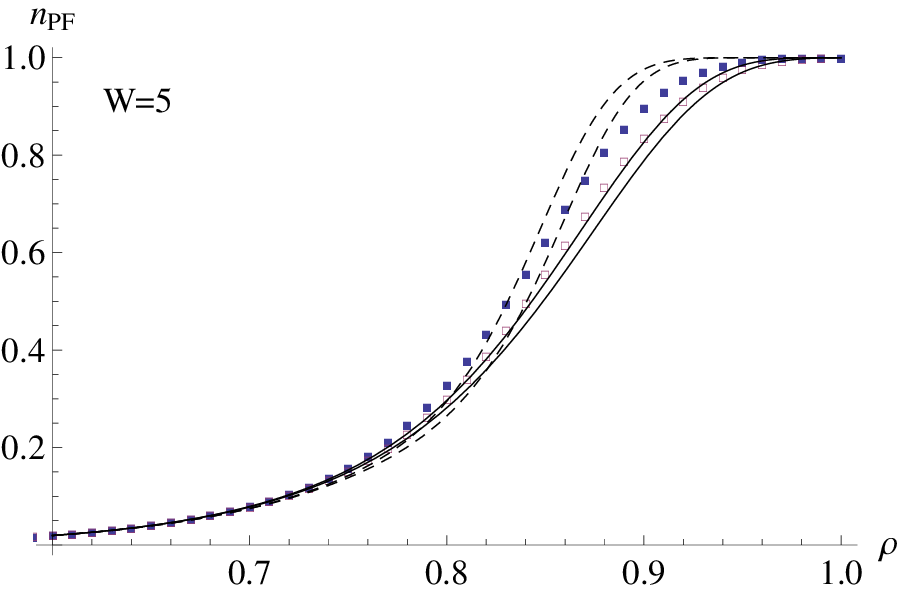}
\includegraphics[width=120pt]{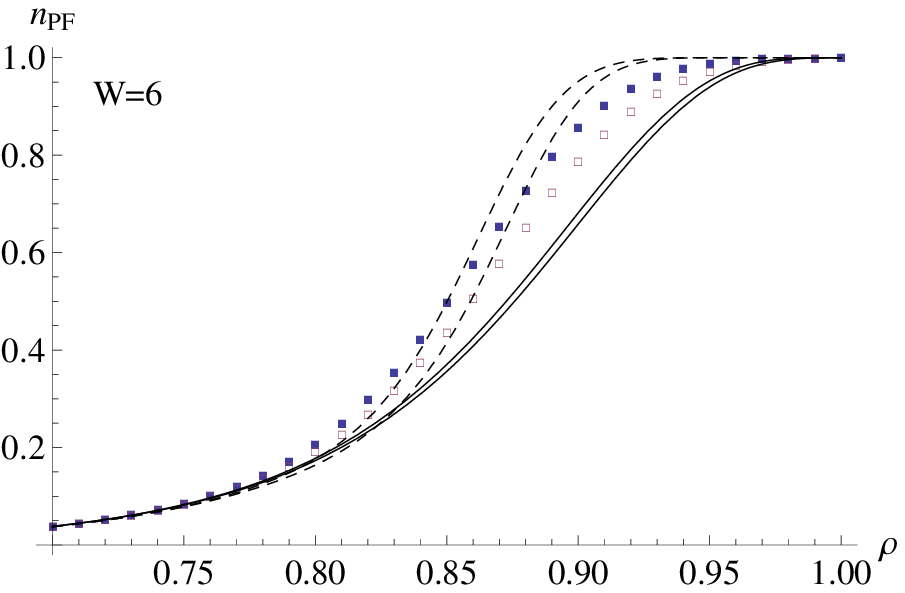}
\caption{
Fraction of frozen particles vs. density for $W=3,4,5,6$ with hard-wall boundaries. Dashed lines are result of the approximation of Section \ref{secdiv} for the KA (higher curve) and FA (lower curve) models. Continuous lines are results from the approximation of Section \ref{sec36} for the KA (higher curve) and FA (lower curve) models. Full (KA) and empty (FA) squares are numerical results. Numerical simulations were done with $L=200W$. Analytical results are for $L=\infty$. As $W$ increases, larger sections should be included. For $W=3,4$, the current approximation is better than the approximation of Section \ref{secdiv}, but for $W\geq5$, longer sections are required.
}
\label{compw36}
\end{figure}

Figure \ref{compw36} shows the fraction of frozen particles obtained by the numerical results, this improved approximation, and the previous approximation. We see that for $W=3$, it is enough to consider  $l=13$, but for the wider systems we need longer sections. Since the fraction of frozen particles in a section decays roughly exponentially with the section's length (see Fig. \ref{fakasec}), we need to consider only $l\approx2\left\langle l\right\rangle$, where $\left\langle l\right\rangle$ is the average section length, see Eq. (\ref{lav}). As the average section length depends on the density, we can take it at the critical density. We find that for such small widths, the average section length is $6.2$ (for $W=3$), $6.53$ ($W=4$), $7.05$ ($W=5$) and $7.6$ (for $W=6$). This explains why $l=13$ is enough for $W=3$, but $l=10$ is not enough for $W=4$. The number of configurations to scan numerically is $2^{2W\left\langle l\right\rangle}=2^{39}\approx5\times10^{11}$ ($W=3$), $2^{52}\approx4\times10^{15}$ ($W=4$), $2^{75}\approx3\times10^{22}$ ($W=5$) and $2^{96}\approx6\times10^{28}$ ($W=6$).

\section{Internal Structures}
\label{secinter}

\begin{figure}
\includegraphics[width=200pt]{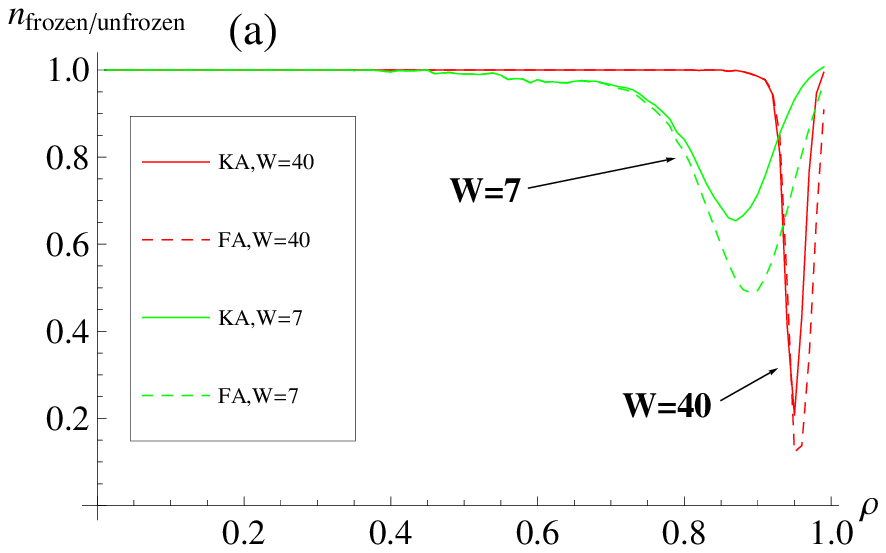}
\includegraphics[width=200pt]{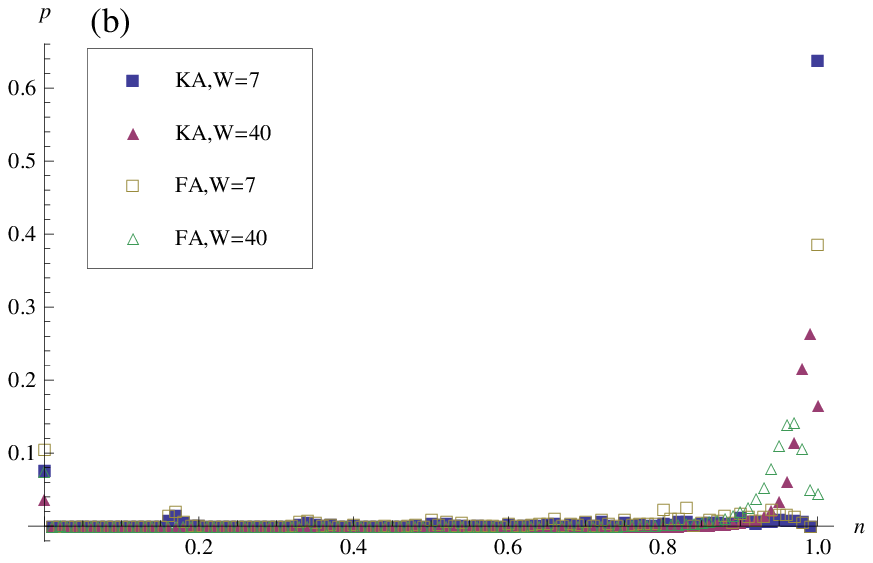}
\caption{(color online) 
(a) Density of sections which are completely frozen or completely unfrozen, hard-wall boundaries. The minimum for $W=40$ is much lower than for $W=7$ because there is a larger chance of a small, inconsequential unfrozen island. (b) The distribution of the fraction of frozen particles per section for $W=7$ and $\rho=0.9$ and for $W=40$ and $\rho=0.95$ (near the minima in panel (a)). Most sections are either completely unfrozen or almost completely frozen.
}
\label{fuf}
\end{figure}

\begin{figure}
\subfigure[$W=100$]{\includegraphics[width=\columnwidth]{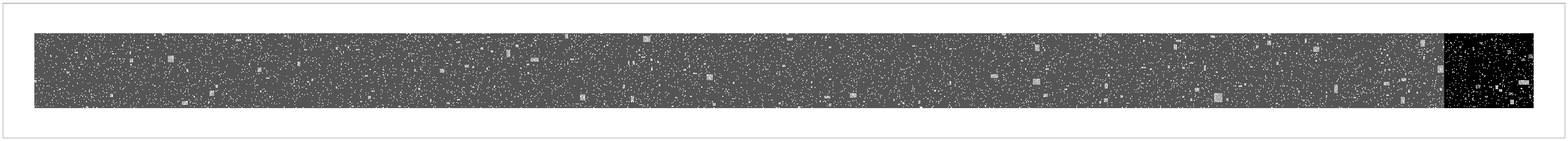}}
\subfigure[$W=7$]{\includegraphics[width=\columnwidth]{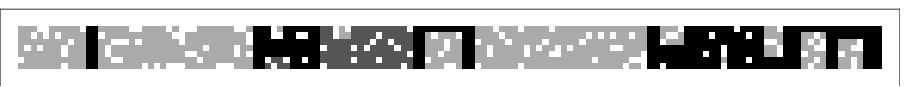}}
\caption{
Snapshots of (a) $100\times2000$ system at $\rho=0.957$, (b) $7\times140$ system at $\rho=0.85$. Both with hard-wall boundaries. Black squares are particles which are frozen in both models, dark gray are frozen only in the KA model, light gray are unfrozen in both models, and white squares are vacancies. (a) contains two sections, which are both almost completely frozen for KA, but only one of them is frozen for FA. Small scattered islands are also visible. In (b) there are $11$ sections, showing many different behaviors. Zooming in on the pictures allows to see the details. In print in the $100\times2000$ system, what appears brightest are the light-gray areas, since the individual white sites are too small to be seen.
}
\label{grid100}
\end{figure}

The main qualitative difference between systems of large widths and small widths is the internal structures within the sections. In wide systems, the vast majority of sections are either almost completely frozen or completely unfrozen, while in narrow systems there is a significant number of sections which are partially frozen. The reason for this difference is the existence of small unfrozen ``islands", which are small regions that do not effect their surroundings. For example, a structure of the form $\begin{array}{cc}10\\00\end{array}$ is unfrozen (in both the KA and the FA model), but it does not necessarily cause the entire section to be unfrozen. When the system is wide these islands are not important, but in a narrow system they are. Figure \ref{fuf} shows the density of completely frozen/unfrozen sections in a wide system ($W=40$) and in a narrow system ($W=7$), and Fig. \ref{grid100} shows a snapshot of the system, highlighting the frozen and unfrozen particles.

\begin{figure}
\includegraphics[width=\columnwidth]{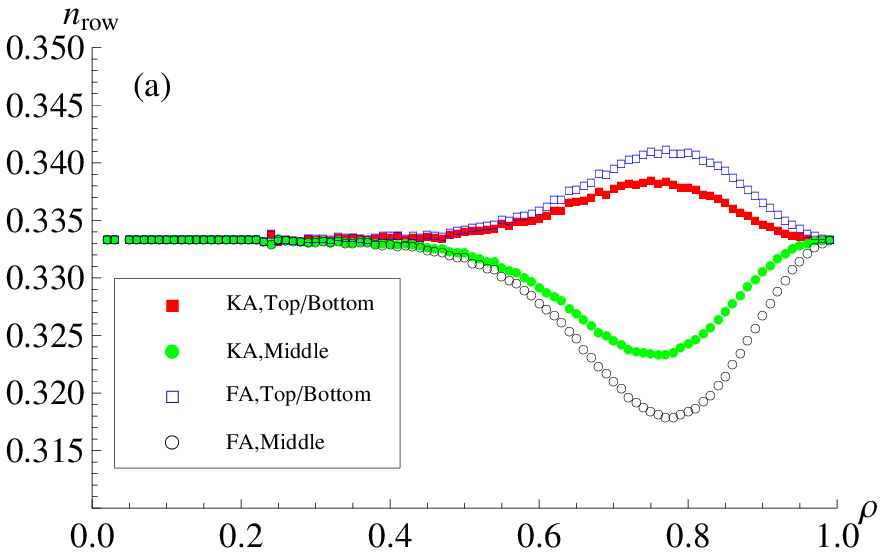}
\includegraphics[width=\columnwidth]{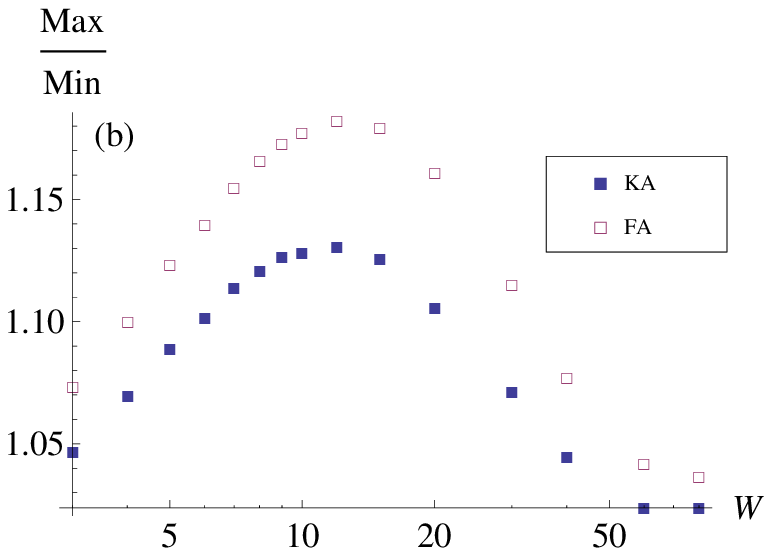}
\caption{(color online) 
(a) Probability that a frozen particle belongs to a certain row at $W=3$. At some width-dependent density, the probability of a frozen particle to be in the extreme rows is maximal. (b) The ratio between the maximum for the extreme rows and the minimum for the middle row(s) as a function of $W$. This is maximal at $W=12$.
}
\label{relrows}
\end{figure}

A quantitative way to measure the effect of the boundary conditions is by noting that with hard-wall boundary conditions, the particles in the top and bottom rows have a slightly higher probability of being frozen than those in the middle rows. This happens because if within a section the top or bottom row is completely full, then all the particles in it are frozen, while for the middle rows this condition is not sufficient. However, as seen from Fig. \ref{relrows}, this difference is very small. Although at first the relative difference between the probability of being frozen at the edges and at the middle grows with $W$, this relative difference reaches a maximum at $W=12$ and then decreases for larger $W$. Hence, we can say that above $W=12$ the boundary conditions become less important. 

\section{Summary}

In this paper we investigated the effect of confinement and shape of container on the jamming transition. We derived an analytical approximation for the fraction of frozen particles in rectangular systems in both the Kob-Andersen and Fredrickson-Andersen kinetically-constrained models, by dividing the system into independent sections and using the notion of critical droplets, which was derived previously for square systems in the FA model. The number of these critical droplets is controlled, in addition to the system's size and particle density, by a single parameter $\lambda$. We showed that the effective value of $\lambda$ does not change much when the system's length increases, and that its value in rectangular systems is approximately the same as in square systems. Also, its value in the Kob-Andersen model is higher than in the Fredrickson-Andersen model, which means that the fraction of frozen particles in the KA model is higher than in the FA model, as expected by comparing the kinetic constraints of the two models. From both the numerical simulations and the analytical expressions, we found that the transition from an unjammed state, where most of the particles are free, to a jammed state, where most of the particles are frozen, occurs over a very narrow range of densities. Using our approximation, we found the critical density at which this transition occurs, $\rho_{c}$, which converges to $1$ as the system size increases. We also found that the width of the transition scales as $(1-\rho_{c})^{2}$.

For infinite tunnels, we derived an exact result for very narrow systems (widths $1$ and $2$) for both the KA model and the FA model. The technique we used can also be applied to wider systems. For infinite systems of width $3-6$ we showed that it is enough to explicitly count the number of frozen particles in sections shorter than twice the typical section length, since the number of frozen particles in a section decays exponentially with the section's length. Also, using the analytical approximation for general rectangles, we found a simple expression relating the critical density with the width of an infinite tunnel, which can be verified in experiments. In particular, we found that the critical density in channels decreases algebraically with the system's width, $1-\rho_{c}\sim 1/\sqrt{W}$, much faster than the logarithmic decrease in square systems, $1-\rho_{c}\sim1/\ln L$. These two different expressions for the critical density show that the jamming transition depends not only on the system size but also on its shape and the relation between the system's width and length.

Our idea of dividing the system into independent sections can also be applied in three-dimensional systems, and will be addressed in future work. It will also be interesting to check the effects of confinement on the behavior of other kinetically-constrained models, such as jamming percolation models \cite{knights,jeng}, and to employ our approach for studying jamming in driven systems, such as 
granular matter flowing  
in a narrow tube.

\begin{acknowledgments}
We thank Svilen Kozhuharov and Vincenzo Vitelli for fruitful discussions, and Yariv Kafri and Peter Sollich for critical reading of the manuscript. This research was supported by the Israel Science Foundation grants No. $617/12$, $1730/12$.
\end{acknowledgments}

\appendix
\begin{widetext}
\section{Calculation of the Fraction of Frozen Particles in Eq. (\ref{npfmain0})}
\label{apnpf}
In this section we calculate explicitly the sums in Eq. (\ref{npfmain0}).
We first note that $Q(n,l)$ is proportional to $\rho^{n}v^{lW-n}$, such that 
\begin{align}
nQ(n,l)=\rho\partial_{\rho}\left[Q(n,l)\right] ,
\end{align}
where $\partial_{\rho}\left[f(\rho)\right]$ is the partial derivative of $f(\rho)$ with respect to $\rho$ while assuming that $v$ is constant. Only after differentiating we use the relation $v+\rho=1$. Hence, $n_{PF}$ can be simplified by
\begin{align}
&n_{PF}=\frac{\sum_{n,l,m}\rho\partial_{\rho}\left[\rho^{mW}\right]Q(n,l)+\rho^{mW}\rho\partial_{\rho}\left[Q(n,l)\right]e^{-kl}}{\rho\partial_{\rho}\left[\sum_{n,l,m}\rho^{mW}Q(n,l)\right]} ,
\end{align}
We can now perform the sums over $n$, such that
\begin{align}
n_{PF}=\frac{\sum_{l,m}\rho\partial_{\rho}\left[\rho^{mW}\right]Q(l)+\rho^{mW}\rho\partial_{\rho}\left[Q(l)\right]e^{-kl}}{\rho\partial_{\rho}\left[\sum_{l,m}\rho^{mW}Q(l)\right]} ,
\end{align}
with $Q(l)=\sum_{n}Q(l,n)$. In order to have the derivative in the nominator outside the sum, we artificially change some of the $\rho$ to $\rho'$, and after differentiating set $\rho'=\rho$, and thus
\begin{align}
n_{PF}=\frac{\rho\partial_{\rho}\left[\sum_{l,m}\rho^{mW}Q(l,\rho')+\rho'^{mW}Q(l,\rho)e^{-kl}\right]}{\rho\partial_{\rho}\left[\sum_{l,m}\rho^{mW}Q(l,\rho)\right]} . \label{npfmain2}
\end{align}
For ease of calculation, we divide $n_{PF}$ into three parts
\begin{align}
n_{PF}=\frac{A_{1}+A_{2}}{B} ,
\end{align}
where
\begin{align}
&A_{1}=\rho\partial_{\rho}\left[\sum_{l,m}\rho^{mW}Q(l,\rho')\right] ,\nonumber\\
&A_{2}=\rho\partial_{\rho}\left[\sum_{l,m}\rho'^{mW}Q(l,\rho)e^{-kl}\right] ,\nonumber\\
&B=\rho\partial_{\rho}\left[\sum_{l,m}\rho^{mW}Q(l,\rho)\right] ,
\end{align}
and calculate each part separately.

\subsection{Calculation of $Q(l)$}
For $l=1$ we denote by $0\leq f<W$ the number of particles in the column, such that
\begin{align}
Q(1)=\sum^{W-1}_{f=0}\rho^{f}v^{W-f}\left(\begin{array}{c}W\\f\end{array}\right)=\left(\rho+v\right)^{W}-\rho^{W}=1-\rho^{W} .\label{ql1}
\end{align}

For the longer sections, we denote by $r_{1}$ and $r_{2}$ the number of particles in the rightmost and leftmost columns, and by $n_{i}$ the number of columns with $i$ particles in them. Thus, the density of sections of length $l$ is
\begin{align}
&Q(l)=\sum^{W-1}_{r_{1}=0}\sum^{W-1}_{r_{2}=0}\sum^{\left\lfloor\frac{l-1}{2}\right\rfloor}_{n_{W}=0}\prod^{W-1}_{i=1}\sum^{l-2-\sum^{W}_{j=i+1}n_{j}}_{n_{i}=0}\rho^{r_{1}+r_{2}}v^{2W-r_{1}-r_{2}}\left(\begin{array}{c}W\\r_{1}\end{array}\right)\left(\begin{array}{c}W\\r_{2}\end{array}\right)\rho^{\sum^{W}_{j=1}jn_{j}}\times\nonumber\\
&\times v^{\sum^{W}_{j=1}\left(W-j\right)n_{j}+W\left(l-2-\sum^{W}_{j=1}n_{j}\right)}\left(\begin{array}{c}l-n_{W}-1\\n_{w}\end{array}\right)\left(\begin{array}{c}l-2-\sum^{W}_{j=i+1}n_{j}\\n_{i}\end{array}\right)\left(\begin{array}{c}W\\i\end{array}\right)^{n_{i}} ,
\end{align}
where $\left\lfloor x\right\rfloor$ is the integer part of $x$. The first binomial in the second line is the number of ways to arrange $n_{W}$ columns among the $l-2$ available places such that there are no two adjacent full columns. The upper limit of the sum over $n_{W}$ is such because above it the binomial is zero. The second binomial in the second line is the number of ways to arrange $n_{i}$ columns among the columns not yet taken by the already-placed columns. Note that we do not sum over $n_{0}$, since it must satisfy $n_{0}=l-2-\sum^{W}_{i=1}n_{i}$. 

Summing over $r_{1}, r_{2}$ yields
\begin{align}
&Q(l)=\left[\left(\rho+v\right)^{W}-\rho^{W}\right]^{2}\sum^{\left\lfloor\frac{l-1}{2}\right\rfloor}_{n_{W}=0}\left(\frac{\rho}{v}\right)^{Wn_{W}}v^{W\left(l-2\right)}\left(\begin{array}{c}l-n_{W}-1\\n_{W}\end{array}\right)\times\nonumber\\
&\times\prod^{W-1}_{i=1}\sum^{l-2-\sum^{W}_{j=i+1}n_{j}}_{n_{i}=0}\left[\left(\frac{\rho}{v}\right)^{i}\left(\begin{array}{c}W\\i\end{array}\right)\right]^{n_{i}}\left(\begin{array}{c}l-2-\sum^{W}_{j=i+1}n_{j}\\n_{i}\end{array}\right) .
\end{align}

We now note that
\begin{align}
\prod^{W-1}_{i=1}\sum^{N-\sum^{W-1}_{j=i+1}n_{j}}_{n_{i}=0}\left[g_{i}\right]^{n_{i}}\left(\begin{array}{c}N-\sum^{W-1}_{j=i+1}n_{j}\\n_{i}\end{array}\right)=\left[\sum^{W-1}_{i=0}g_{i}\right]^{N} ,\label{proof1}
\end{align}
where $g_{0}=1$. The proof for this equation is given in Appendix \ref{approof1}. We can now write
\begin{align}
&Q(l)=\left[\left(\rho+v\right)^{W}-\rho^{W}\right]^{2}\sum^{\left\lfloor\frac{l-1}{2}\right\rfloor}_{n_{W}=0}\left(\frac{\rho}{v}\right)^{Wn_{W}}v^{W\left(l-2\right)}\left(\begin{array}{c}l-n_{W}-1\\n_{W}\end{array}\right)\left[\sum^{W-1}_{i=0}\left(\frac{\rho}{v}\right)^{i}\left(\begin{array}{c}W\\i\end{array}\right)\right]^{l-2-n_{W}}=\nonumber\\
&=\left[\left(\rho+v\right)^{W}-\rho^{W}\right]^{2}\sum^{\left\lfloor\frac{l-1}{2}\right\rfloor}_{n_{W}=0}\left[\frac{1}{\left(1+\frac{v}{\rho}\right)^{W}-1}\right]^{n_{W}}\left(\begin{array}{c}l-n_{W}-1\\n_{W}\end{array}\right) .
\end{align}
Performing the sum over $n_{W}$ yields
\begin{align}
&Q(l)=\frac{\left[\left(\rho+v\right)^{W}-\rho^{W}\right]^{l}\left[\left(1+\frac{v}{\rho}\right)^{W}-1\right]}{\sqrt{\left[\left(1+\frac{v}{\rho}\right)^{W}+1\right]^{2}-4}}\times\nonumber\\
&\times\left[\left(\frac{\left(1+\frac{v}{\rho}\right)^{W}-1+\sqrt{\left[\left(1+\frac{v}{\rho}\right)^{W}+1\right]^{2}-4}}{2\left[\left(1+\frac{v}{\rho}\right)^{W}-1\right]}\right)^{l}-\left(\frac{\left(1+\frac{v}{\rho}\right)^{W}-1-\sqrt{\left[\left(1+\frac{v}{\rho}\right)^{W}+1\right]^{2}-4}}{2\left[\left(1+\frac{v}{\rho}\right)^{W}-1\right]}\right)^{l}\right]=\nonumber\\
&=\frac{1-\rho^{W}}{\sqrt{1+2\rho^{W}-3\rho^{2W}}}\left[x^{l}_{+}-x^{l}_{-}\right] ,\label{ql}
\end{align}
where
\begin{align}
x_{\pm}=\frac{1-\rho^{W}\pm\sqrt{\left(1+\rho^{W}\right)^{2}-4\rho^{2W}}}{2} .
\end{align}
Only in the last step did we use $v+\rho=1$. Using $l=1$ in the general equation for $l>1$ yields the same result we found in Eq. (\ref{ql1}) for $Q(1)$. Note also that $Q(l)$ depends on $\rho$ and $W$ only via $\rho^{W}$. An interesting point to make is that
\begin{align}
Q(2)=Q(3)=\left(1-\rho^{W}\right)^{2} .
\end{align}
Also, we find that $Q(l+1)\geq Q(l)$ for all values of $\rho^{W}$, and equality holds only for $l=2$.

\subsection{The Final Result for $n_{PF}$}
Using Eq. (\ref{ql}) we can now calculate the sums in Eq. (\ref{npfmain2}). All the sums are such that $l+m\leq L$. Also, $m$
is greater or equal to $2$ except for the case $l=L$ where $m=0$.

\subsubsection{The Denominator}
The denominator of $n_{PF}$ in Eq. (\ref{npfmain2}) is
\begin{align}
&B=\rho\partial_{\rho}\left[\sum_{l,m}\rho^{mW}Q(l,\rho)\right]=\rho\partial_{\rho}\left[Q(L,\rho)+\rho^{WL}+\sum^{L-2}_{l=1}\sum^{L-l}_{m=2}\rho^{mW}Q(l,\rho)\right] .
\end{align}
Performing the sum over $m$ yields
\begin{align}
&B=\rho\partial_{\rho}\left[Q(L,\rho)+\rho^{WL}+\sum^{L-2}_{l=1}\frac{\rho^{2W}-\rho^{W\left(L-l+1\right)}}{1-\rho^{W}}Q(l,\rho)\right] .
\end{align}
Performing the sum over $l$, differentiating with respect to $\rho$ and finally setting $v=1-\rho$ yields
\begin{align}
&B=\frac{W\rho^{1-2W}}{1-\rho^{W}}+\frac{W\rho^{W\left(L+1\right)}\left[2-2\rho^{W}-\rho^{W+1}+L\left(2-5\rho^{W}+3\rho^{2W}\right)\right]}{\left(1-\rho^{W}\right)\left(2-3\rho^{W}\right)^{2}}+LW\rho^{WL}-\nonumber\\
&-\frac{W\left[x^{L}_{+}+x^{L}_{-}\right]}{2\rho^{2W}\left(1+3\rho^{W}\right)\left(2-3\rho^{W}\right)^{2}}\times\nonumber\\
&\times\left[2L\rho^{2W}\left(2-3\rho^{W}\right)\left(1-\rho^{W}-3\rho^{2W}+3\rho^{W+1}\right)+\left(1+3\rho^{W}\right)\left(4\rho+2\rho^{3W}-8\rho^{W+1}+\rho^{2W+1}+\rho^{3W+1}\right)\right]+\nonumber\\
&+\frac{\sqrt{\left(1+\rho^{W}\right)^{2}-4\rho^{2W}}W\left[x^{L}_{+}-x^{L}_{-}\right]}{2\rho^{2W}\left(1-\rho^{W}\right)\left(2+3\rho^{W}-9\rho^{2W}\right)^{2}}\left[2\left(L+1\right)\rho^{2W}\left(2+3\rho^{W}-9\rho^{2W}\right)\left(1-3\rho^{W}+3\rho^{2W}-\rho^{W+1}\right)-\right.\nonumber\\
&\left.-\left(1-\rho^{W}\right)\left(4\rho+4\rho^{2W}+12\rho^{3W}-36\rho^{4W}+54\rho^{5W}+12\rho^{W+1}-15\rho^{2W+1}-54\rho^{3W+1}+27\rho^{4W+1}\right)\right] .\label{denfinal}
\end{align}

\subsubsection{First Part of the Nominator}
The first term in the nominator of Eq. (\ref{npfmain2}) is
\begin{align}
&A_{1}=\rho\partial_{\rho}\left[\sum_{l,m}\rho^{mW}Q(l,\rho')\right]=\rho\partial_{\rho}\left[Q(L,\rho')+\rho^{WL}+\sum^{L-2}_{l=1}\sum^{L-l}_{m=2}\rho^{mW}Q(l,\rho')\right]=\nonumber\\
&=\rho\partial_{\rho}\left[\rho^{WL}+\sum^{L-2}_{l=1}\sum^{L-l}_{m=2}\rho^{mW}Q(l,\rho')\right] ,
\end{align}
where in the last transition we note that $\partial_{\rho}Q(l,\rho')=0$, since here $Q$ is a function of $\rho'$ and not $\rho$. Summing over $m$ yields
\begin{align}
A_{1}=\rho\partial_{\rho}\left[\rho^{WL}+\sum^{L-2}_{l=1}\frac{\rho^{2W}-\rho^{\left(L-l+1\right)W}}{1-\rho^{W}}Q(l,\rho')\right] .
\end{align}
Summing over $l$, differentiating with respect to $\rho$ (not $\rho'$), and lastly setting $v=1-\rho$ and $\rho'=\rho$ yields
\begin{align}
&A_{1}=\frac{W\left(2-\rho^{W}\right)}{1-\rho^{W}}+\frac{W\rho^{\left(L+1\right)W}\left[L\left(2-5\rho^{W}+3\rho^{2W}\right)+3-4\rho^{W}\right]}{\left(1-\rho^{W}\right)\left(2-3\rho^{W}\right)^{2}}+WL\rho^{WL}-\nonumber\\
&-\frac{W\left[x^{L}_{+}+x^{L}_{-}\right]\left(1-\rho^{W}\right)\left(8-9\rho^{W}\right)}{2\left(2-3\rho^{W}\right)^{2}}-\frac{W\left[x^{L}_{+}-x^{L}_{-}\right]\left(1-\rho^{W}\right)\left(8-7\rho^{W}-3\rho^{2W}\right)}{2\left(2-3\rho^{W}\right)^{2}\sqrt{1+2\rho^{W}-3\rho^{2W}}} .\label{nom1final}
\end{align}

\subsubsection{Second Part of the Nominator}
The second part of the nominator is
\begin{align}
&A_{2}=\rho\partial_{\rho}\left[\sum_{l,m}\rho'^{mW}Q(l,\rho)e^{-lk}\right]=\rho\partial_{\rho}\left[Q(L,\rho)e^{-Lk}+\sum^{L-2}_{l=1}\sum^{L-l}_{m=2}\rho'^{mW}Q(l,\rho)e^{-lk}\right]=\nonumber\\
&=\rho\partial_{\rho}\left[Q(L,\rho)e^{-Lk}+\sum^{L-2}_{l=1}\frac{\rho'^{2W}-\rho'^{\left(L-l+1\right)W}}{1-\rho'^{W}}Q(l,\rho)e^{-lk}\right] .
\end{align}
Summing over $l$, differentiating with respect to $\rho$ and lastly setting $v=1-\rho$ and $\rho'=\rho$ yields
\begin{align}
&Nom2=\frac{We^{k}\rho^{2W}\left[e^{2k}\left(\rho-\rho^{W}\right)+\rho^{W}\left(1-\rho^{W}\right)^{2}\right]y^{2}_{1}}{1-\rho^{W}}-\frac{We^{k}\rho^{W\left(L+1\right)}\left[\left(1-\rho^{W}\right)^{2}+e^{2k}\rho^{W}\left(\rho-\rho^{W}\right)\right]y^{2}_{2}}{1-\rho^{W}}-\nonumber\\
&-\frac{e^{-kL}WL\left(1-\rho\right)\left[x^{L}_{+}+x^{L}_{-}\right]}{2\left(1+3\rho^{W}\right)}+\frac{e^{-kL}\rho^{W}W\left[x^{L}_{+}+x^{L}_{-}\right]}{2\left(1-\rho^{W}\right)^{2}\left(1+3\rho^{W}\right)}\times\nonumber\\
&\times\left\{e^{3k}\left(1-\rho^{W}\right)\left[\left(1+3\rho^{W}\right)\left(1-4\rho^{W}+4\rho^{2W}-\rho^{W+1}\right)-L\left(1-\rho^{W}-4\rho^{2W}+6\rho^{3W}+\rho^{W+1}-3\rho^{2W+1}\right)\right]\times\right.\nonumber\\
&\left.\times\left[y^{2}_{1}-y^{2}_{2}\right]+e^{2k}\left(1-\rho^{W}\right)^{2}\left[2+4\rho^{W}-6\rho^{2W}-L\left(1+\rho^{W}-3\rho^{2W}+\rho^{W+1}\right)\right]\left[\rho^{W}y^{2}_{1}-y^{2}_{2}\right]-\right.\nonumber\\
&\left.-e^{4k}\left(1-\rho^{W}\right)\left[2+4\rho^{W}-6\rho^{2W}-L\left(2+\rho^{W}-6\rho^{2W}+3\rho^{W+1}\right)\right]\left[y^{2}_{1}-\rho^{W}y^{2}_{2}\right]+\right.\nonumber\\
&\left.+e^{5k}\left[1+2\rho^{W}-3\rho^{2W}-L\left(1-3\rho^{2W}+2\rho^{W+1}\right)\right]\left[y^{2}_{1}-\rho^{2W}y^{2}_{2}\right]\right\}-\nonumber\\
&-\frac{e^{-kL}W\left[x^{L}_{+}-x^{L}_{-}\right]\left[4\rho^{W}\left(1-\rho\right)-L\left(1+\rho-2\rho^{W}\right)\left(1+3\rho^{W}\right)\right]}{2\left(1+3\rho^{W}\right)\sqrt{1+2\rho^{W}-3\rho^{2W}}}+\nonumber\\
&+\frac{e^{-kL}W\rho^{W}\left[x^{L}_{+}-x^{L}_{-}\right]}{2\left(1-\rho^{W}\right)\left(1+3\rho^{W}\right)\sqrt{1+2\rho^{W}-3\rho^{2W}}}\times\nonumber\\
&\times\left\{-e^{2k}\left(1-\rho^{W}\right)\left[2+6\rho^{W}-6\rho^{2W}-6\rho^{3W}+4\rho^{2W+1}-L\left(1+3\rho^{W}\right)\left(1-\rho^{W}-\rho^{2W}+\rho^{W+1}\right)\right]\left[\rho^{W}y^{2}_{1}-y^{2}_{2}\right]-\right.\nonumber\\
&\left.-e^{3k}\left(1-\rho^{W}\right)\left[1+\rho^{W}-10\rho^{2W}+6\rho^{3W}-\rho^{W+1}+3\rho^{2W+1}-L\left(1+3\rho^{W}\right)\left(1-2\rho^{W}+\rho^{W+1}\right)\right]\left[y^{2}_{1}-y^{2}_{2}\right]+\right.\nonumber\\
&\left.+e^{4k}\left[2+6\rho^{W}-10\rho^{2W}+6\rho^{3W}-4\rho^{W+1}-L\left(1+3\rho^{W}\right)\left(2-3\rho^{W}+\rho^{W+1}\right)\right]\left[y^{2}_{1}-\rho^{W}y^{2}_{2}\right]-\right.\nonumber\\
&\left.-e^{5k}\left[1+4\rho^{W}-3\rho^{2W}-2\rho^{W+1}-L\left(1+2\rho^{W}-3\rho^{2W}\right)\right]\left[y^{2}_{1}-\rho^{2W}y^{2}_{2}\right]\right\} , \label{nom2final}
\end{align}
where
\begin{align}
&y_{1}=\frac{1}{\left(e^{k}+\rho^{W}\right)\left(1-\rho^{W}\right)-e^{2k}} ,\nonumber\\
&y_{2}=\frac{1}{\left(e^{k}+1\right)\left(1-\rho^{W}\right)-e^{2k}\rho^{W}} .
\end{align}

\section{Proof for Eq. (\ref{proof1})}
\label{approof1}
Here we prove that
\begin{align}
\prod^{W-1}_{i=1}\sum^{N-\sum^{W-1}_{j=i+1}n_{j}}_{n_{i}=0}
\left[g_{i}\right]^{n_{i}}\left(\begin{array}{c}N-\sum^{W-1}_{j=i+1}n_{j}\\n_{i}\end{array}\right)=\left[1+\sum^{W-1}_{i=1}g_{i}\right]^{N} .
\end{align}
We do this by induction on $W$. For $W=2$ it holds because
\begin{align}
\sum^{N-0}_{n_{1}=0}\left[g_{1}\right]^{n_{1}}\left(\begin{array}{c}N-0\\n_{1}\end{array}\right)=\left[1+g_{1}\right]^{N} .
\end{align}

For $W+1$ we first sum over $n_{1}$
\begin{align}
&\prod^{W}_{i=1}\sum^{N-\sum^{W}_{j=i+1}n_{j}}_{n_{i}=0}
\left[g_{i}\right]^{n_{i}}\left(\begin{array}{c}N-\sum^{W}_{j=i+1}n_{j}\\n_{i}\end{array}\right)=\prod^{W}_{i=2}\sum^{N-\sum^{W}_{j=i+1}n_{j}}_{n_{i}=0}
\left[g_{i}\right]^{n_{i}}\left(\begin{array}{c}N-\sum^{W}_{j=i+1}n_{j}\\n_{i}\end{array}\right)\left[1+g_{1}\right]^{N-\sum^{W}_{j=2}n_{j}}=\nonumber\\
&=\left[1+g_{1}\right]^{N}\prod^{W}_{i=2}\sum^{N-\sum^{W}_{j=i+1}n_{j}}_{n_{i}=0}
\left[\frac{g_{i}}{1+g_{1}}\right]^{n_{i}}\left(\begin{array}{c}N-\sum^{W}_{j=i+1}n_{j}\\n_{i}\end{array}\right) .
\end{align}
We now define $h_{i}$ such that
\begin{align}
h_{i}=\frac{g_{i+1}}{1+g_{1}} ,
\end{align}
and rewrite the sum as
\begin{align}
&\prod^{W}_{i=1}\sum^{N-\sum^{W}_{j=i+1}n_{j}}_{n_{i}=0}
\left[g_{i}\right]^{n_{i}}\left(\begin{array}{c}N-\sum^{W}_{j=i+1}n_{j}\\n_{i}\end{array}\right)=\left[1+g_{1}\right]^{N}\prod^{W-1}_{i=1}\sum^{N-\sum^{W-1}_{j=i+1}n_{j}}_{n_{i}=0}
\left[h_{i}\right]^{n_{i}}\left(\begin{array}{c}N-\sum^{W-1}_{j=i+1}n_{j}\\n_{i}\end{array}\right) .
\end{align}
Since the sum is now only up to $W-1$ we know what it is
\begin{align}
&\prod^{W}_{i=1}\sum^{N-\sum^{W}_{j=i+1}n_{j}}_{n_{i}=0}
\left[g_{i}\right]^{n_{i}}\left(\begin{array}{c}N-\sum^{W}_{j=i+1}n_{j}\\n_{i}\end{array}\right)=\left[1+g_{1}\right]^{N}\left[1+\sum^{W-1}_{i=1}h_{i}\right]^{N}=\left[1+\sum^{W}_{i=1}g_{i}\right]^{N} ,
\end{align}
as required.

\section{Derivation of Eq. (\ref{ka2end})}
\label{apka2}
The sum in Eq. (\ref{w2eq1}) over each subsection is identical and independent of the others, thus we can transform Eq. (\ref{w2eq1}) to
\begin{align}
&N^{KA,hw}_{PF}(\rho,2)=2\rho v\sum^{\infty}_{d=0}\left(\rho^{3}v\right)^{d}\left[\left(3d+1\right)\left(\sum_{t=z,l}\sum^{\infty}_{h=0}\sum^{h-\delta_{t,r}}_{f=0}v^{h}\rho^{h+2f}\left(\begin{array}{c}h\\f\end{array}\right)\right)^{d+1}+\right.\nonumber\\
&\left.+\left(d+1\right)\left(\sum_{t=z,r}\sum^{\infty}_{h=0}\sum^{h-\delta_{t,r}}_{f=0}v^{h}\rho^{h+2f}\left(\begin{array}{c}h\\f\end{array}\right)\right)^{d}\left(\sum_{t=z,r}\sum^{\infty}_{h=0}\sum^{h-\delta_{t,r}}_{f=0}v^{h}\rho^{h+2f}\left(\begin{array}{c}h\\f\end{array}\right)\left(h+f+f\delta_{t,z}\right)\right)\right] .
\end{align}
Calculating the sums over $f,h$ and $t$ yields
\begin{align}
&N^{KA,hw}_{PF}(\rho,2)=2\rho v\sum^{\infty}_{d=0}\left(\rho^{3}v\right)^{d}\left[\left(3d+1\right)C^{d+1}_{2}\left(\rho\right)+\left(d+1\right)C^{d}_{2}\left(\rho\right)C_{1}\left(\rho\right)\right] ,\nonumber\\
&C_{1}\left(\rho\right)=\frac{\rho v\left[2+3\rho^{2}-6\rho^{5}v-2\rho^{4}v^{2}-2\rho^{6}v^{2}+3\rho^{8}v^{2}\right]}{\left(1-\rho^{3}v\right)^{2}\left[1-\rho v\left(1+\rho^{2}\right)\right]^{2}} ,\nonumber\\
&C_{2}\left(\rho\right)=\frac{1+\rho v\left(1-\rho^{2}\right)}{\left(1-\rho^{3}v\right)\left[1-\rho v\left(1+\rho^{2}\right)\right]} .
\end{align}
Summing over $d$ yields Eq. (\ref{ka2end})

\section{Derivation of Eq. (\ref{fa2begin})}
\label{apfa2}
We rewrite some of the $\rho$ in Eq. (\ref{fa2begin}) as $\rho'$, so that $N^{FA}_{PF}$ can be simplified to
\begin{align}
&N^{FA,hw}_{PF}(\rho,2)=2\sum^{\infty}_{d=0}\prod^{d+1}_{i=1}\sum^{\infty}_{h_{i}=0}\sum^{h_{i}}_{f_{i}=0}v^{d}v^{h_{i}}\rho^{f_{i}}\rho'^{3d+h_{i}+f_{i}+1}\left(\begin{array}{c}h_{i}\\f_{i}\end{array}\right) v\left[3d+\sum^{d+1}_{j=1}\left(h_{j}+f_{j}\right)+1\right]=\nonumber\\
&=2\rho'\partial_{\rho'}\sum^{\infty}_{d=0}\prod^{d+1}_{i=1}\sum^{\infty}_{h_{i}=0}\sum^{h_{i}}_{f_{i}=0}v^{d}v^{h_{i}}\rho^{f_{i}}\rho'^{3d+h_{i}+f_{i}+1}\left(\begin{array}{c}h_{i}\\f_{i}\end{array}\right)v .\label{eqd1}
\end{align}
As before, since each of the $d+1$ sums over $h_{i}$ and $f_{i}$ are independent, we can write $N^{FA}_{PF}$ as
\begin{align}
N^{FA,hw}_{PF}(\rho,2)=2\rho'\partial_{\rho'}\sum^{\infty}_{d=0}v^{d+1}\rho'^{3d+1}\left[\sum^{\infty}_{h=0}\sum^{h}_{f=0}v^{h}\rho^{f}\rho'^{h+f}\left(\begin{array}{c}h\\f\end{array}\right)\right]^{d+1} .
\end{align}
Calculating the sum over $h$ and $f$ yields
\begin{align}
N^{FA,hw}_{PF}(\rho,2)=2\rho'\partial_{\rho'}\sum^{\infty}_{d=0}v^{d+1}\rho'^{3d+1}\left[\frac{1}{1-v\rho'\left(1+\rho\rho'\right)}\right]^{d+1} .\label{eqd2}
\end{align}
Calculating the sum over $d$, differentiating with respect to $\rho'$, and finally setting $\rho'=\rho$, yields
\begin{align}
N^{FA,hw}_{PF}(\rho,2)=\frac{2\rho v\left(1+3\rho^{3}v\right)}{\left[1-\rho v\left(1+2\rho^{2}\right)\right]^{2}} .
\end{align}

\section{Derivation of Eq. (\ref{ka2per})}
\label{apkap2}
We rewrite Eq. (\ref{ka2per}) as
\begin{align}
N^{KA,per}_{PF}(\rho,2)=2\rho\partial_{\rho}\sum^{\infty}_{d=0}\left(\rho^{3}v\right)^{d}\prod^{d+1}_{i=1}\sum^{\infty}_{h_{i}=0}\sum^{h_{i}}_{f_{i}=0}v^{h_{i}}\rho^{h_{i}+2f_{i}}\left(\begin{array}{c}h_{i}\\f_{i}\end{array}\right)\rho v .
\end{align}
This is exactly Eq. (\ref{eqd1}) with $\rho'=\rho$, and therefore we can use Eq. (\ref{eqd2})
\begin{align}
N^{KA,per}_{PF}(\rho,2)=2\rho\partial_{\rho}\sum^{\infty}_{d=0}v^{d+1}\rho^{3d+1}\left[\frac{1}{1-v\rho\left(1+\rho^{2}\right)}\right]^{d+1} .
\end{align}
Summing over $d$ and differentiating with respect to $\rho$ yields
\begin{align}
N^{KA,per}_{PF}(\rho,2)=\frac{2\rho v\left(1+4\rho^{3}v\right)}{\left[1-\rho v\left(1+2\rho^{2}\right)\right]^{2}} .
\end{align}

\end{widetext}


\begin{thebibliography}{0}
\expandafter\ifx\csname natexlab\endcsname\relax\def\natexlab#1{#1}\fi
\expandafter\ifx\csname bibnamefont\endcsname\relax
  \def\bibnamefont#1{#1}\fi
\expandafter\ifx\csname bibfnamefont\endcsname\relax
  \def\bibfnamefont#1{#1}\fi
\expandafter\ifx\csname citenamefont\endcsname\relax
  \def\citenamefont#1{#1}\fi
\expandafter\ifx\csname url\endcsname\relax
  \def\url#1{\texttt{#1}}\fi
\expandafter\ifx\csname urlprefix\endcsname\relax\def\urlprefix{URL }\fi
\providecommand{\bibinfo}[2]{#2}
\providecommand{\eprint}[2][]{\url{#2}}

\end{thebibliography}


\begin{thebibliography}{999}

\bibitem{jamming}{A. J. Liu and S. R. Nigel, \textit{Nature}, \textbf{396}, 21 (1998)}

\bibitem{industry}{http://www.rocksystems.com/machinery/conveyors}

\bibitem{industry2}{http://www.slb.com/services/drilling/cementing/\\equipment/cement\_slurry\_defoamer.aspx}

\bibitem{nature}{R. A. Bagnold, \textit{Geological Survey Professional Paper, 422-I}, \textbf{I-20} (1966)}

\bibitem{nature2}{http://vulcan.wr.usgs.gov/Glossary/LavaTubes/\\framework.html} 

\bibitem{durian}{D. J. Durian, \textit{Phys. Rev. E}, \textbf{55}, 1739 (1997)}

\bibitem{haxton}{T. K. Haxton and A. J. Liu, \textit{Europhys. Lett.}, \textbf{90}, 66004 (2010)}

\bibitem{ohern}{C. S. O'Hern, L. E. Silbert, A. J. Liu, and S. R. Nagel, \textit{Phys. Rev. E}, \textbf{68}, 011306 (2003)}

\bibitem{ningxu}{N. Xu, V. Vitelli, M. Wyart, A. J. Liu, and S. R. Nagel, \textit{Phys. Rev. Lett}, \textbf{102}, 038001 (2009)}

\bibitem{lerner}{E. Lerner, I. Procaccia, and J. Zylberg, \textit{Phys. Rev. Lett.}, \textbf{102}, 125701 (2009)}

\bibitem{barrat}{B. Andreotti, J.-L. Barrat, and C. Heussinger, \textit{Phys. Rev. Lett.}, \textbf{109}, 105901 (2012)}

\bibitem{exp1}{N. Saklayen, G. L. Hunter, K. V. Edmond, and E. R. Weeks, arXiv:1209.1108v1}

\bibitem{exp2}{A. I. Campbell and M. D. Haw, \textit{Soft Matter}, \textbf{6}, 4688 (2010).}

\bibitem{exp3}{K. N. Nordstrom, E. Verneuil, P. E. Arratia, A. Basu, Z. Zhang, A. G. Yodh, J. P. Gollub, and D. J. Durian, \textit{Phys. Rev. Lett.}, \textbf{105}, 175701 (2010).} 

\bibitem{exp4}{M. A. Lohr, A. M. Alsayed, B. G. Chen, Z. Zhang, R. D. Kamien, and A. G. Yodh, \textit{Phys. Rev. E}, \textbf{81}, 040401(R) (2010)}

\bibitem{behringer1}{K. E. Daniels and R. P. Behringer, \textit{J. Stat. Mech.}, P07018 (2006)}

\bibitem{behringer2}{D. Bi, J. Zheng, B. Chakraborty, and R. P. Behringer, \textit{Nature}, \textbf{480}, 355 (2011)}

\bibitem{review}{F. Ritort and P. Sollich, \textit{Advances in Physics}, \textbf{52}, 219 (2003)}

\bibitem{review2}{J. P. Garrahan, P. Sollich, and C. Toninelli, \textit{Dynamical Heterogeneities in Glasses, Colloids, and Granular Media}, edited by L. Berthier, G. Biroli, J.-P. Bouchaud, L. Cipelletti, and W. van Saarloos (Oxford University Press 2011), Chap. 10; arXiv:1009.6113v1 (2010)}

\bibitem{kronig}{A. Kronig and J. Jackle, \textit{J. Phys.: Condens. Matter}, \textbf{6}, 7633 (1994)}

\bibitem{fieldings}{S. M. Fielding, \textit{Phys. Rev. E}, \textbf{66}, 016103 (2002)}

\bibitem{toninelli}{C. Toninelli, G. Biroli, D. S. Fisher, \textit{Phys. Rev. Lett.}, \textbf{92}, 185504 (2004)}

\bibitem{sellitto}{M. Sellitto, G. Biroli, and C. Toninelli, \textit{Europhys. Lett.}, \textbf{69}(4), 496 (2005)}

\bibitem{knights}{C. Toninelli, G. Biroli, and D. S. Fisher, \textit{Phys. Rev. Lett.}, \textbf{96}, 035702 (2006)}

\bibitem{DFOT}{J. P. Garrahan, R. L. Jack, V. Lecomte, E. Pitard, K. van Duijvendijk, and F. van Wijland, \textit{Phys. Rev. Lett.}, \textbf{98}, 195702 (2007)}

\bibitem{sellitto2}{M. Sellitto, \textit{Phys. Rev. Lett.}, \textbf{101}, 048301 (2008)}

\bibitem{spiral}{F. Corberi and L. F. Cugliandolo, \textit{J. Stat. Mech.}, P09015 (2009)}

\bibitem{jeng}{M. Jeng and J. M. Schwarz, \textit{Phys. Rev. E}, \textbf{81}, 011134 (2010)}

\bibitem{shokef}{Y. Shokef and A. J. Liu, \textit{Euro. Phys. Lett.}, \textbf{90}, 26005 (2010)}

\bibitem{elmatad}{Y. S. Elmatad, R. L. Jack, D. Chandler, and J. P. Garrahan, \textit{Proc. Natl. Acad. Sci. USA}, \textbf{107}, 12793 (2010)}

\bibitem{driving}{F. Turci, E. Pitard, and M. Sellitto, \textit{Phys. Rev. E}, \textbf{86}, 031112 (2012)}

\bibitem{kipnis}{C. Kipnis, C. Marchioro, and E. Presutti, \textit{J. Stat. Phys}, \textbf{27}, 65 (1980)}

\bibitem{derrida2}{B. Derrida, \textit{Phys. Rev. Lett.}, \textbf{45}, 79 (1980)}

\bibitem{derrida}{B. Derrida, \textit{Phys. Rev. B}, \textbf{24}, 2613 (1981)}

\bibitem{bouchaud}{J. -P. Bouchaud, \textit{J. Phys. I France}, \textbf{2}, 1705 (1992)}

\bibitem{asep}{B. Derrida, M. R. Evans, and D. Mukamel, \textit{J. Phys. A: Math. Gen.}, \textbf{26}, 4911 (1993)}

\bibitem{monthus}{C. Monthus and J. -P. Bouchaud, \textit{J. Phys. A: Math. Gen.}, \textbf{29}, 3847 (1996)}

\bibitem{nemeth}{Z. T. Nemeth and H. Lowen, \textit{Phys. Rev. E}, \textbf{59}, 6824 (1999)}

\bibitem{scheidler}{P. Scheidler, W. Kob, and K. Binder, \textit{Europhys. Lett.}, \textbf{52}(3), 277 (2000)}

\bibitem{varnik}{F. Varnik, J. Baschnagel, and K. Binder, \textit{Phys. Rev. E}, \textbf{65}, 021507 (2002)}

\bibitem{srebro}{Y. Srebro and D. Levine, \textit{Phys. Rev. Lett}, \textbf{93}, 240601 (2004)}

\bibitem{zrp}{M. R. Evans and T. Hanney, \textit{J. Phys. A: Math. Gen.}, \textbf{38}, R195 (2005)}

\bibitem{lang}{S. Lang, V. Botan, M. Oettel, D. Hajnal, T. Franosch, and R. Schilling, \textit{Phys. Rev. Lett.}, \textbf{105}, 125701 (2010)}

\bibitem{kamodel}{W. Kob and H.C. Andersen, \textit{Phys. Rev. E}, \textbf{48}, 4364 (1993)}

\bibitem{famodel}{G. H. Fredrickson and H.C. Andersen, \textit{Phys. Rev. Lett}, \textbf{53}, 1244 (1984)}

\bibitem{adler}{J. Adler, \textit{Physica A}, \textbf{171}, 453 (1991)}

\bibitem{bp}{A. E. Holroyd, \textit{Probability Theory and Related Fields}, \textbf{125}, 195 (2003)}

\bibitem{aharony}{J. Adler, D. Stauffer, and A. Aharony, \textit{J. Phys. A: Math. Gen.}, \textbf{22}, L297 (1989)}

\bibitem{holroydslow}{J. Gravner and A. E. Holroyd, \textit{The Annals of Applied Probability}, \textbf{18}, 909 (2008)}

\bibitem{olsson}{P. Olsson and S. Teitel, \textit{Phys. Rev. Lett}, \textbf{99}, 178001 (2007)}

\end{thebibliography}
\end{document}